\documentclass[journal, a4paper, 10pt]{IEEEtran}

\ifCLASSINFOpdf
\else
\fi

\ifCLASSOPTIONcompsoc
  \usepackage[caption=false,font=normalsize,labelfont=sf,textfont=sf]{subfig}
\else
  \usepackage[caption=false,font=footnotesize]{subfig}
\fi

\usepackage[T1]{fontenc}
\usepackage{amsmath,amsfonts}
\usepackage{algorithm}
\usepackage{algpseudocode}
\usepackage{array}
\usepackage{makecell}
\usepackage{adjustbox}
\usepackage{tabularx}
\usepackage{amssymb}
\usepackage[a4paper,right=0.62in, left=0.62in, top=1.92cm, bottom=4.4cm]{geometry}
\usepackage{amsmath} 
\usepackage{xcolor}
\usepackage{textcomp}
\usepackage{stfloats}
\usepackage{amsmath}
\usepackage{url}
\usepackage{verbatim}
\usepackage{graphicx,booktabs,multirow,bm}
\usepackage{subfig}
\usepackage{xcolor}
\usepackage{cite}
\usepackage{colortbl}

\newcommand{\h}[1]{{#1}}
\newcommand{\g}[1]{{#1}}

\newcommand{\x}{\boldsymbol{x}}
\newcommand{\z}{\boldsymbol{z}}
\newcommand{\y}{\boldsymbol{y}}

\newcommand{\hz}{\hat{\boldsymbol{z}}}
\newcommand{\hy}{\hat{\boldsymbol{y}}}
\newcommand{\hZ}{\hat{Z}}

\newcommand{\bE}{\mathbb{E}}

\allowdisplaybreaks[4]
\DeclareMathOperator*{\argmin}{arg\,min}
\begin{document}

\title{Tackling Distribution Shifts in Task-Oriented Communication with Information Bottleneck}
\author{\IEEEauthorblockN{Hongru~Li, Jiawei~Shao, Hengtao~He, Shenghui~Song,\\ Jun Zhang,~\textit{Fellow,~IEEE,} and {Khaled B.~Letaief,~\textit{Fellow,~IEEE}}}


\thanks{This article was presented in part at the IEEE Global Communication Conference (GLOBECOM), in Dec. 2023 \cite{li2023task}}
\thanks{The authors are with the Department of Electronic and Computer Engineering, The Hong Kong University of Science and Technology (HKUST), Hong Kong (e-mail: hlidm@connect.ust.hk, jiawei.shao@connect.ust.hk, eehthe@ust.hk, eeshsong@ust.hk, eejzhang@ust.hk, eekhaled@ust.hk). (The corresponding author is Hengtao He.)}}



\maketitle

\pagestyle{empty}  
\thispagestyle{empty}
\begin{abstract}

Task-oriented communication aims to extract and transmit task-relevant information to significantly reduce the communication overhead and transmission latency.
However, the \textit{unpredictable} distribution shifts between training and test data, including \textit{domain shift} and \textit{semantic shift}, can dramatically undermine the system performance.
In order to tackle these challenges, it is crucial to ensure that the encoded features can generalize to \textit{domain-shifted} data and detect \textit{semantic-shifted} data, while remaining compact for transmission.
In this paper, we propose a novel approach based on the information bottleneck (IB) principle and invariant risk minimization (IRM) framework.
The proposed method aims to extract compact and informative features that possess high capability for effective \textit{domain-shift generalization} and accurate \textit{semantic-shift detection} without any knowledge of the test data during training.
Specifically, we propose an invariant feature encoding approach based on the IB principle and IRM framework for \textit{domain-shift} generalization, which aims to find the causal relationship between the input data and task result by minimizing the complexity and domain dependence of the encoded feature.
Furthermore, we enhance the task-oriented communication with the label-dependent feature encoding approach for \textit{semantic-shift detection} which achieves joint gains in IB optimization and detection performance. To avoid the intractable computation of the IB-based objective, we leverage variational approximation to derive a tractable upper bound for optimization.
Extensive simulation results on image classification tasks demonstrate that the proposed scheme outperforms state-of-the-art approaches and achieves a better rate-distortion tradeoff.

\end{abstract}


\vspace{0.2cm}
\section{Introduction}

As the global deployment of 5G wireless networks accelerates, the potential development of 6G has been widely explored~\cite{letaief2019roadmap}. This transition is significantly driven by the groundbreaking advancements in artificial intelligence (AI) technologies. 
Traditional communication systems have been significantly enhanced by AI tools~\cite{o2017introduction} in many aspects, including signal detection~\cite{he2020model}, channel estimation~\cite{he2022beamspace,yu2023adaptive}, source/channel coding~\cite{jssc_deniz}, and resource allocation~\cite{shen2020graph}. However, the rapid evolution of AI technologies also introduces unprecedented challenges~\cite{zhu2020toward,shi2020communication}. \g{These challenges encompass the need for extensive data transmission capacities and the demand for ultra-low latency, underscoring the crucial need for developing innovative techniques to efficiently meet these emerging demands.}

\g{A promising technique to address these challenges is task-oriented communication, which contrasts sharply with traditional communication systems. Traditional data-oriented communication transmits a huge volume of raw data (e.g., images, high-definition videos) without considering the semantic meaning of the information, which incurs high communication overhead and latency. In contrast, task-oriented communication~\cite{vfe} aims to distill and transmit compact and informative representations of data for downstream tasks.}
It extracts and transmits task-relevant information while ignoring redundant task-irrelevant information, thereby significantly reducing the communication overhead and achieving low transmission latency. Specifically, task-relevant information comes from the \textit{causal} components of data, e.g. digits in the Colored-MNIST dataset, representing the semantic information in the data. Conversely, task-irrelevant information comes from the non-causal components of data, e.g., colors in the Colored-MNIST dataset, and is independent of the semantic information of data.

\g{Despite its promising potential for efficient information transmission, task-oriented communication introduces new design challenges. One of the critical challenges is the mismatch between the training and test data distributions, which can dramatically undermine the system performance. To address this issue, in this paper, we propose an information bottleneck (IB)-based approach which can extract compact and informative features with higher generalizability and distinguishability.}

\vspace{0.2cm}
\subsection{Related Works}
By conceiving information transmission as a data reconstruction task and utilizing an autoencoder architecture~\cite{autoencoders}, deep learning (DL) has been seamlessly integrated into the communication systems, leading to the development of DL-based joint source-channel coding (DeepJSCC)~\cite{jssc_deniz,jankowski2020wireless} for image transmission. The training of DeepJSCC is in an end-to-end manner, where the transmitter and receiver are jointly optimized and the wireless channel is treated as a non-learnable component.
DeepJSCC directly maps raw data into channel codewords at the transmitter and recovers the data from the received channel codewords at the receiver.
Besides image data, DeepJSCC has been extensively studied across various modalities, including text-based~\cite{yao2022semantic}, video-based~\cite{jiang2022wireless}, speech-based~\cite{weng2021semantic}, and multimodal-based~\cite{zhang2024unified,xie2022task} information transmission. These studies consistently demonstrate performance improvements over traditional approaches with separate source-channel coding approaches. 

\h{However,  the above research is still \textit{data-oriented} and focuses on the high-fidelity data reconstruction at the receiver side, leading to high communication overhead.}
\g{To this end, \textit{task-oriented communication} is a promising candidate to significantly reduce the communication overhead by transmitting only task-relevant information.}
Engaging results have been achieved in applying this method to various scenarios. In~\cite{iccshao}, the authors proposed a model splitting method to jointly design on-device computation and communication for edge inference. 
Instead of optimizing the transceivers based on data reconstruction metrics (e.g., mean square error (MSE) or bit error rate), task-oriented communication directly optimizes the transceiver based on the inference performance, without data recovery at the receiver. 
This idea is further combined with the IB principle for edge inference~\cite{vfe}. The IB principle aims to maximize inference performance while minimizing communication overhead~\cite{tishby2000informationIB}, aligning with the objective of task-oriented communication. Moreover, the distributed IB (DIB) method~\cite{aguerri2019distributed} has been extended to multi-device collaborative task-oriented communication~\cite{shao2022task}, where the coded redundancy is efficiently minimized among multiple devices.
A recent work~\cite{shao2023task} leveraged IB principle and temporal entropy modeling techniques to reduce temporal redundancy for multi-camera video analysis tasks and achieved a better rate-performance tradeoff on sequential data processing. 
Meanwhile, the authors in~\cite{xie2022robust_IB} proposed a robust IB-based method that formulates the informativeness-robustness tradeoff, depicting the relationship between the coded redundancy and the inference performance, which enhances the robustness of task-oriented communication.

Although task-orineted communication has achieved great success~\cite{xie2024deep}, one critical challenge, i.e., the distribution shifts between the training and test data, is overlooked. Such shifts incur a substantial degradation to the reliability and security of task-oriented communications. Recent research has developed several schemes for tackling this challenge.  Specifically, the authors of~\cite{xu2021wireless,wu2023vision} proposed to use the attention modules~\cite{vaswani2017attention} to combat the channel distribution shifts, which enhances the robustness of task-oriented communication. However, the attention-based approaches suffer from expensive memory consumption, which is not acceptable in practical systems. To avoid this issue, a hypernetwork-based method~\cite{xie2024deep} was proposed to generate parameters for both the transmitter and receiver. This method outperforms the attention-based method and can be seamlessly integrated into existing task-oriented communication systems with minimal memory overhead. To handle the distribution shift of input data, a recent work~\cite{dai2023toward} proposed a method to update transceiver parameters for each test instance by leveraging the overfitting property of neural networks. Nevertheless, it incurs significant computation and communication overhead due to the transmission of gradient information. Furthermore, a task-oriented communication system based on cycleGAN~\cite{zhang2022deep} was developed to transform unseen data into similar known data which can handle the dynamic data environment in task-oriented communication. However, it requires test data for online adaptation, which is impractical in real-world scenarios. Additionally, as discussed in~\cite{mescheder2018training}, GAN-based methods suffer from convergence issue due to the nature of the min-max optimization problem.

Despite the aforementioned efforts to address the data distribution shifts in task-oriented communication, there are still three critical challenges. Firstly, the unpredictability of the test dataset prior to deployment renders the retraining methods~\cite{zhang2022deep} ineffective. Additionally, the cost of parameter updates in retraining is significantly high due to the gradient information exchanged between transceivers~\cite{dai2023toward}. Secondly, it is difficult to collect a substantial amount of target domain data in the test stage without reliable detection techniques. \h{Finally, existing methods assume that the test data share the same label space with the training data, disregarding the possibility of novel classes not being present during training. Thus, there is an urgent call for developing a retraining-free scheme which can detect or generalize to unknown data to address the distribution shifts problem.}

\subsection{Contributions}
In this paper, we propose a novel method based on the IB principle and invariant risk minimization (IRM) approach to tackle the distribution shifts in task-oriented communication. Our main contributions are summarized as follows
\begin{itemize} 
    \item 
    As far as our knowledge extends, we are the first to confront the formidable challenge of distribution shifts in task-oriented communication, all while operating with no prior knowledge of test data.
    Our method aims to extract compact and informative features with high generalizability and distinguishability for effective \textit{domain-shift} generalization and \textit{semantic-shift} detection without compromising the rate-distrotion tradeoff. Furthermore, the proposed method eliminates the need for accessing to test domain data or test-time \h{update}.
    \item We propose a novel invariant principle enhanced IB method, named variational invariant feature encoding (VIFE), which can effectively extract \textit{task-relevant information} and achieves \textit{domain-shift generalization} in task-oriented communication. Our proposed method combines the benefits of the IB principle with IRM techniques. \h{With empirical examples and analysis, we demonstrate that the proposed method can simultaneously achieve low communication overhead and high generalization performance.}
    \item 
    \h{To solve the unknown data issue during deployment}, we further propose the variational label-dependent feature encoding (VLFE) schemes to efficiently detect \textit{semantic-shift} data.
    Specifically, we incorporate the label conditional latent prior into the IB objective and utilize the contrastive learning technique. It can ensure the separation of the latent space associated with each in-distribution data class while retaining low communication overhead.
    \item To verify the effectiveness of our proposed scheme, we conduct evaluations on two image classification tasks. Extensive simulation results demonstrate that the proposed method significantly outperforms the baselines in terms of both \textit{domain-shit} generalization and \textit{semantic-shift} detection performance, while maintaining a favorable rate-distortion tradeoff.
\end{itemize}

\subsection{Organization and Notations}
The rest of this paper is organized as follows. Section \ref{sec-model_problem} presents the system model and formulates the problem.
Section \ref{sec-iib} investigates the IFE method for enhancing \textit{domain-shift} generalization and analyzes the joint gain in communication overhead and generalization performance. 
Section \ref{sec-ccib} proposes the LFE objective for efficient \textit{semantic-shift} detection. The extensive simulations are presented in Section \ref{sec-exp}. Finally, Section \ref{sec-conclusion} concludes this paper.

In this paper, upper-case letters (e.g. $X,Y$) represent random variables, and lower-case letters (e.g. $\boldsymbol{x},\boldsymbol{y}$) represent the realizations of the corresponding random variables. $H(X)$ and $H(X|Y)$ are the entropy of $X$ and conditional entropy of $X$ given $Y$. $I(X;Y)$ and $I(X;Y|Z)$ represent the mutual information between $X$ and $Y$ and the conditional mutual information between $X$ and $Y$ given $Z$. $D_{KL}(p||q)$ stands for Kullback-Leibler divergence between two probability distribution $p(\boldsymbol{x})$ and $q(\boldsymbol{x})$. The expectation of $X$ is denoted as $\mathbb{E}(X)$. We further denote $X~\sim N(\boldsymbol{\mu},\boldsymbol{\Sigma})$ as $X$ follows the Gaussian distribution with mean $\boldsymbol{\mu}$ and covariance matrix $\boldsymbol{\Sigma}$. We use $X\sim\text{Bernoulli}{(q)}$ to denote that $X$ follows the Bernoulli distribution with $p(X=1)=q$. Finally, we use $\oplus$ to represent XOR or modulo-two-sum calculation.

\section{System model and problem formulation}\label{sec-model_problem}
\subsection{System Model}\label{subsec_model}

\begin{figure*}[t]
    \centering
    \includegraphics[width=\linewidth]{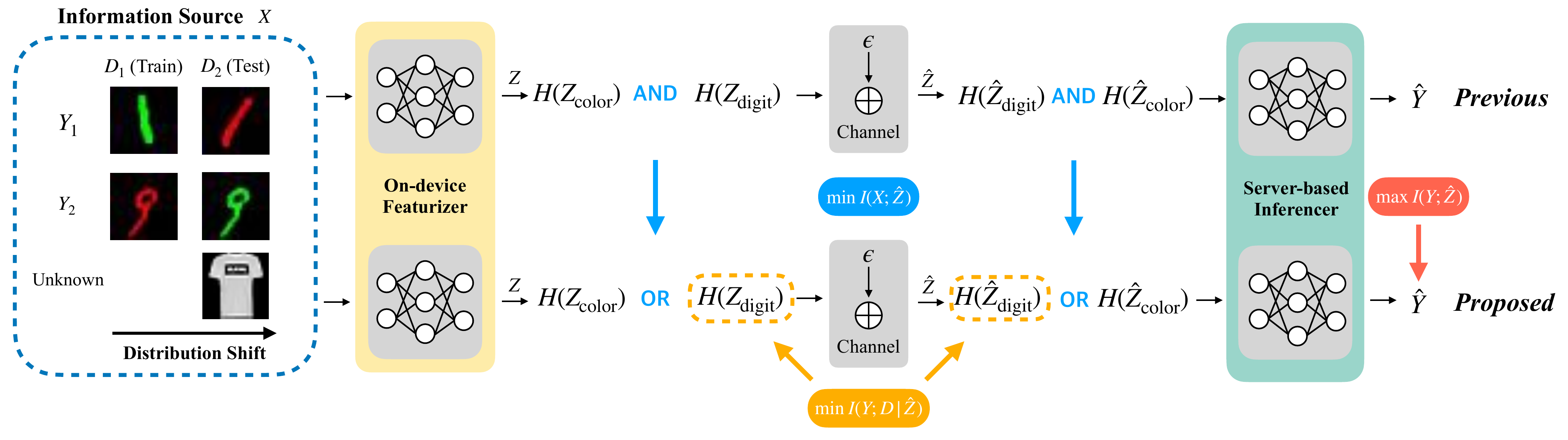}
        \caption{\h{The system model and comparison with previous schemes.
        The information source $X$ includes task-relevant information (e.g., digit), task-irrelevant information (e.g., color) and unknown data during testing. On-device featurizers extract and transmit $z$, while the server-based inferencers derive the task result $\hat{y}$ from the received feature $\hat{z}$. 
        Previous non-causal-aware methods may easily rely on the task-irrelevant features. In contrast, the proposed method aims to discover the causal relationship between $X$ and $Y$ by disregarding the unstable task-irrelevant information. This is achieved by minimizing the complexity ($\min I(X;\hat{Z})$) and domain dependence ($\min I(Y;D|\hat{Z})$) of $\hat{Z}$, and thus the compact and causal features can be selected from input data $X$.}}
    \label{fig:sys}
    \end{figure*}

\h{As shown in Fig.\,\ref{fig:sys}, we consider a task-oriented communication system with a low-end device and an edge server in an \textit{open-world} setting.} In this context, \textit{open-world} denotes an environment where the test data distribution during deployment potentially diverges from that \h{in the training stage}. \h{We assume that the device is resource-constrained and not able to afford the inference of powerful deep neural networks. As a result, it needs the help of the server for inference. To reduce the communication overhead,} the device only extracts task-relevant features from the input data and transmits them to the edge server.

\begin{figure}[t]  
    \centering         
    {\includegraphics[width=0.45\textwidth]{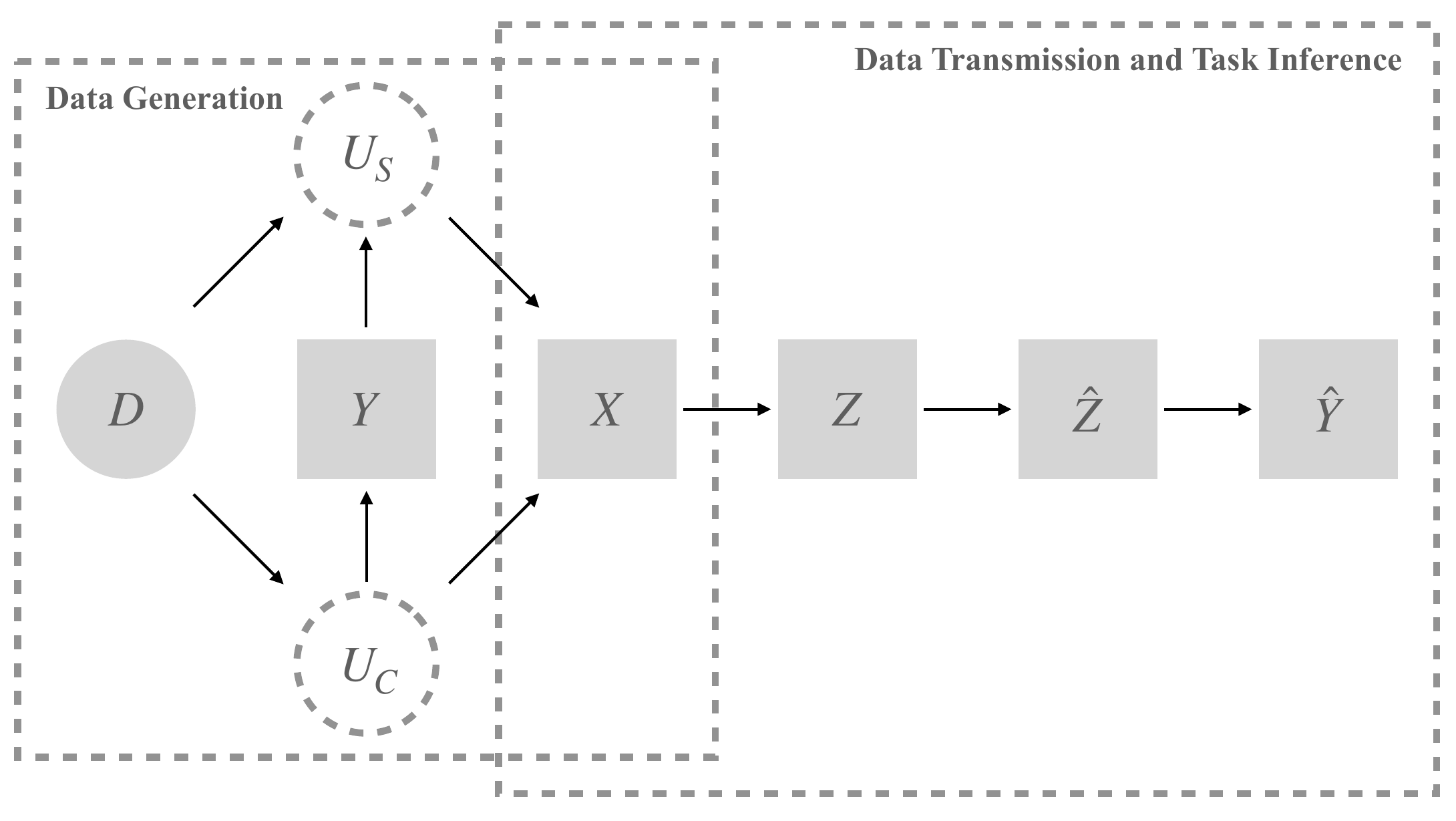}}
    \caption{The probabilistic graphical model of data generation, data transmission and task inference of the proposed task-oriented communication system.}   
    \label{fig:pgm1}         
  \end{figure}

In Fig.\,\ref{fig:pgm1}, we illustrate the probabilistic graphical models (PGM) of the data generation and information transmission process.
Specifically, in the PGM of data generation, we denote the input data sample of the edge device and its corresponding target of the task (e.g. label) as $X$ and $Y$, respectively, with joint distribution $p(\boldsymbol{x},\boldsymbol{y})$. 

The data $X$ consists of two components: 1) the \textit{causal} features $U_C$ (e.g., the digit shape in the Colored MNIST dataset), which directly causes the targets and formulates the task-relevant information; 2) the \textit{spurious} features $U_S$ (e.g., the color in the Colored MNIST dataset), which may have high correlation with targets but does not have any \textit{causal} relationship with them, thereby constituting the task-irrelevant information.
The distribution of $U_S$ may change across different domains $(D)$ but the underlying \textit{causal} relationship between $U_C$ and $Y$ keeps invariant. Moreover, according to the $d$-separation principle\cite{pearl2009causality} and the PGM in Fig.\,\ref{fig:pgm1}, the conditional mutual information $I(D,Y|\cdot)$ should satisfy
\begin{equation}
\begin{aligned}
    I(D;Y|U_C)&=0,
    \label{eq:mi_1}
\end{aligned}
\end{equation}
indicating that given the causal features $U_C$, the label $Y$ is independent of the domain $D$, and 
\begin{equation}
    \begin{aligned}
    I(D;Y|U_C,U_S)&> 0,\\
     I(D;Y|U_S)&> 0,
     \label{eq:mi_2}
    \end{aligned}
\end{equation}
indicating that given spurious features $U_S$, the label $Y$ is dependent on the domain $D$. \eqref{eq:mi_1} comes from $U_C\rightarrow Y$, i.e., $U_C$ is the only parent of $Y$, and \eqref{eq:mi_2} come from $D\rightarrow U_S\leftarrow Y$, i.e., $U_S$ is in the collider path between $D$ and $Y$~\cite{pearl2009causality}.

In the PGM of data transmission and task inference, the on-device feature extractor $\theta$ with the JSCC module maps the input data $X$ into channel codeword $Z$, while the server-based inference model $\phi$ maps the received feature $\hat{Z}$ to the inference result $\hat{Y}$. Thus we can use conditional distribution to model the transmission and inference processes as follows:
\begin{equation}
    p(\x,\hy,\z,\hz) = p(\x)p_\theta(\z|\x)p_{\text{channel}}(\hz|\z)p_\phi(\hy|\hz).
 \end{equation}

To account for the limited transmission power of edge device, we consider an individual transmit power constraint for each feature dimension, ensuring that the power of each feature dimension is below $P_{max}$, i.e., $||z_i||^2\leq P_{max}$.
Without loss of generality\footnote{Note that such a channel model can be easily extended to other communication models by estimating the channel transfer function\cite{vfe, deep_learning_based_ota}.}, we consider an additive white Gaussian noise (AWGN) channel, i.e., $\hz=\z+\epsilon$, $\epsilon\sim{N}({0,\sigma^2\boldsymbol{I}})$. Furthermore, the channel condition can be described by the following peak signal-to-noise ratio (PSNR) metric,
\begin{equation}
    \text{PSNR}=10\log_{10}\frac{P_{max}}{\sigma^2}\text{(dB)}.
    \label{eq:PSNR}
\end{equation}

\vspace{-0.5cm}
\subsection{Problem Description}
The main objective of designing task-oriented communication is to extract \textit{minimal} and \textit{sufficient} features to minimize the communication overhead and inference distortion. To achieve this, we aim to minimize some distortion measurements $d(\y,\hy)$ (e.g. MSE and cross-entropy loss) between the inference result $\hy$ and the ground truth $\y$, while ensuring the communication overhead is below a certain threshold $R$. The solution can be derived by solving the following rate-distortion optimization problem,
\begin{equation}\label{eq:rd}
\begin{aligned}
    \min_{\theta,\phi}\ \ & \bE[d(\y,\hy)] \\
    \text{s.t.}\ \ & I(\hat{Z},X)  \leq R.
\end{aligned}
\end{equation}
Here, we constrain $I(\hat{Z},X)$ instead of $I(Z,X)$ as the coded redundancy can be formulated as $I(\hat{Z},Z)-I(\hat{Z},X)$~\cite{xie2022robust_IB}. By constraining $I(\hat{Z},X)$, task-oriented communication is able to enjoy highly redundant channel coding to counter the dynamic communication environment.

\begin{figure}[t]  
    \centering         
    \subfloat[Natural dataset]{\includegraphics[width=\linewidth]{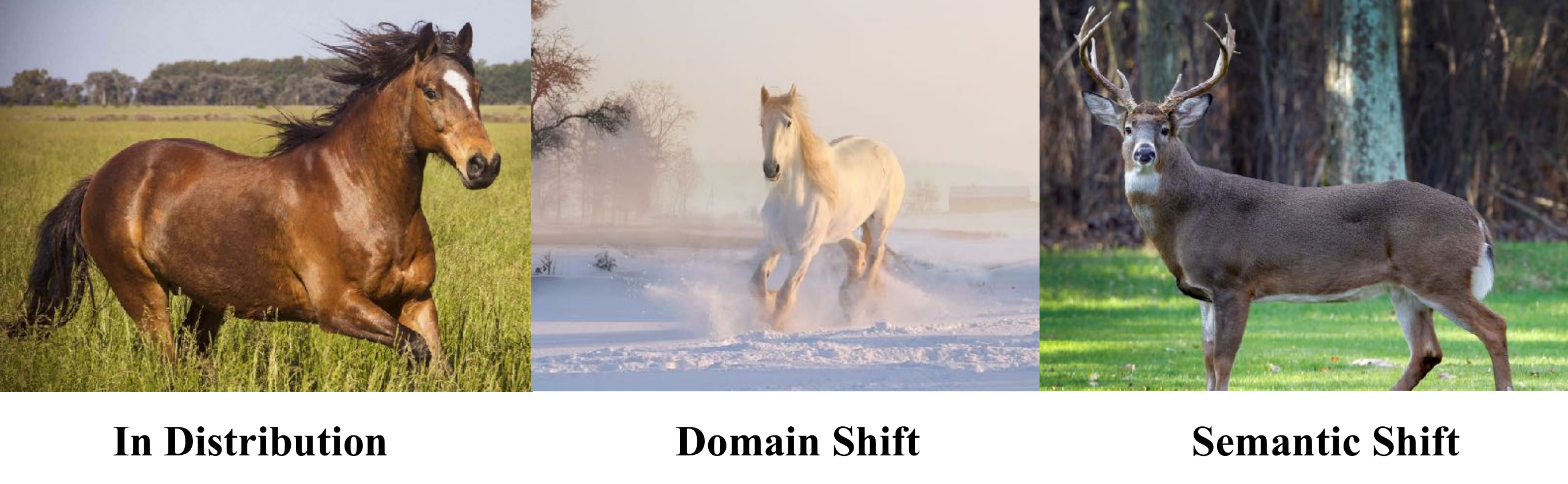}}\label{fig:ood_nature}
    \subfloat[MNIST-related dataset]{\includegraphics[width=\linewidth]{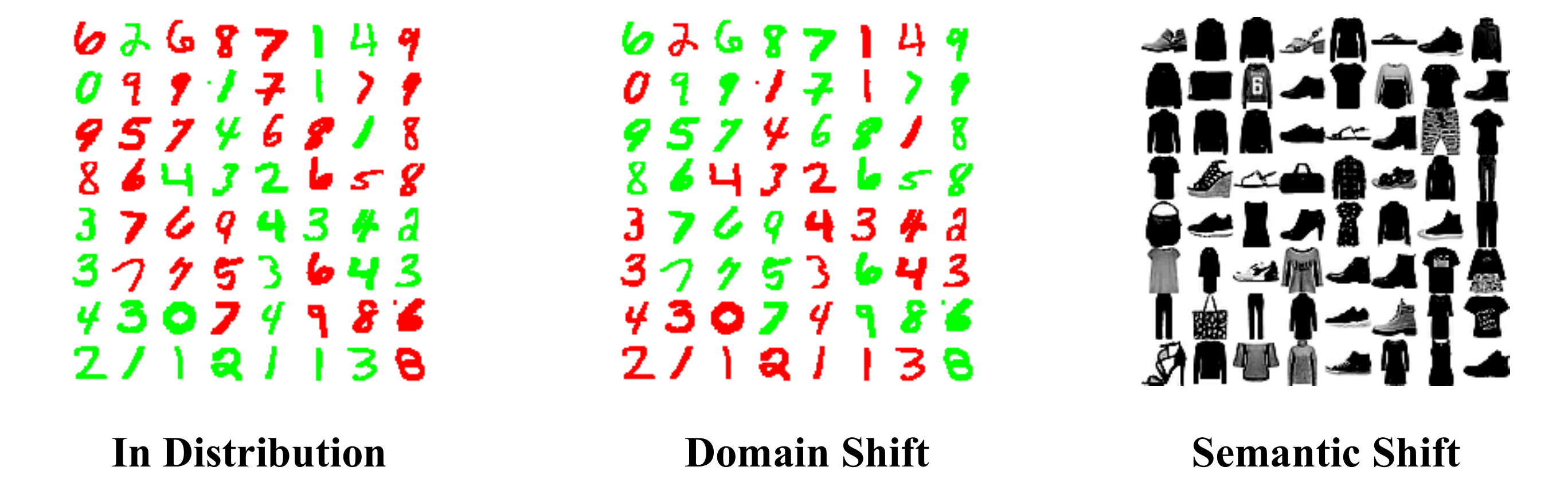}}\label{fig:ood_syn}
    \caption{\h{Examples of distribution shifts} on natural dataset and MNIST-related dataset when the task-oriented communication systems are deployed under the \textit{open-world} assumption. (a) ID data: a horse on grass, domain-shifted data: a horse on snowfield and semantic-shifted data: a deer (b) ID data: gray MNIST dataset, domain-shifted data: Colored-MNIST in which color has a high correlation with digits, and semantic-shifted data: Fashion-MNIST dataset.}
    \label{fig:ood_illurstration}         
  \end{figure}

In practice, the solution of \eqref{eq:rd} is obtained by data-driven DL techniques. Consequently, the obtained task-oriented communication systems may suffer serious performance loss due to data distribution shifts. To illustrate the main challenges for addressing the distribution shifts, we present the taxonomy of three types of data during deployment on natural and MNIST-related datasets in Fig.\,\ref{fig:ood_illurstration}.

\h{The first column in Fig.\,\ref{fig:ood_illurstration} represents the in-distribution (ID) data, 
which is foreseen and trained on by task-oriented communication systems. }
The second column represents the domain-shifted data, which cannot be foreseen before deployment. In this case, the distribution of task-relevant information (i.e., object/digit in Fig.\,\ref{fig:ood_illurstration}) remains unchanged, whereas the distribution of task-irrelevant information (i.e., background and digit color in Fig.\,\ref{fig:ood_illurstration}) has shifted. The third column represents the semantic-shifted data, which has a different label space from the ID data. DL-based models tend to exhibit overconfidence when encountering such data samples~\cite{yang2021generalized}, leading to unreliable results.
Given the distribution shifts in task-oriented communication, an important yet unexplored question arises:

\textit{\textbf{How can a task-oriented communication system be designed to generalize to domain-shifted data and detect semantic shifts without compromising the rate-distortion tradeoff?}}

\section{Variational Invariant Feature Encoding for domain-shift generalization}\label{sec-iib}

In this section, we will introduce an IB principle to address the \textit{domain-shift} generalization challenge in task-oriented communication. We will begin by examining a foundational remote regression example to demonstrate that the vanilla IB method can facilitate domain-shift generalization and reduce communication overhead by constraining the complexity of the extracted features. 
However, it is insufficient to address \textit{domain-shift} generalization challenge as spurious features may exhibit lower complexity compared to causal features.
Thus we further enhance IB with the IRM principle, which encourages the extracted features to have low domain dependence. Finally, to solve the intractable high-dimensional integral in mutual information calculation, we leverage a variational approximation approach to derive a tractable upper bound. We refer our proposed method as variational invariant feature encoding (VIFE).

\subsection{Information Bottleneck for Domain-Shift Generalization}

As shown by the PGM in Fig.\,\ref{fig:pgm1}, the data $X$ contains the causal features $U_C$ and the spurious features $U_S$.
The relation between data and the corresponding label is established by the conditional distribution $P(Y|U_C)$. Furthermore, the label is independent of both the spurious feature distribution $P(U_S)$ and the conditional distribution $P(Y|U_S)$.
{Therefore, it is crucial for the edge device to extract and transmit $U_C$ while disregarding $U_S$. If task-oriented communication effectively learns the causal relationship between $U_C$ and $Y$, it can achieve robust generalization across various domains.}


To illustrate this point, similar to~\cite{arjovsky2019invariant}, we consider a remote regression example over AWGN channel. We assume the data generation process is 
\begin{equation}
    \begin{aligned}
        U_C \leftarrow N(0,\sigma^2_{d_1})\\
        Y \leftarrow U_C + N(0,\sigma^2)\\
        U_S \leftarrow N(0,\sigma^2_{d_2})
        \label{eq:data_gen}
    \end{aligned}
\end{equation}
where $U_C$ and $U_S$ are causal feature and spurious feature, which is consistent with the PGM in Fig.\,\ref{fig:pgm1}; $\sigma_{d_1}$ and $\sigma_{d_2}$ are the data distribution parameters which are domain-specific and $\sigma$ is the label noise. The on-device featurizer extracts features $U_C, U_S$ and transmits them over AWGN channel. The least-squares regressor \eqref{eq:regression} will leverage the received features \eqref{eq:regression_noise} to predict $Y$, satisfying 
\begin{align}
    \hat{Y}&=w_1\hat{U}_C+w_2\hat{U}_S,
    \label{eq:regression}\\
    \hat{U}_C=U_C+\eta_1, \hat{U}_S=&U_S+\eta_2,\eta_1,\eta_2\sim{N}(0,\sigma^2_c).
    \label{eq:regression_noise}
\end{align}

The coefficients $\boldsymbol{w}=(w_1, w_2)$ are derived by solving $\partial \mathcal{L}_{\text{mse}}/\partial \boldsymbol{w}=0$ and the solutions of different conditions are listed below\footnote{The detailed derivation for this part can be found in Appendix~\ref{appendix_example}.},
\begin{itemize}
    \item If the channel condition is good, i.e., $\sigma_c\approx 0$, and we only regress on $U_C$, then we have $(w_1, w_2)=(1,0)$, which is independent of domain-specific parameters, leading to success in generalizing to novel domains.
    \item If the channel condition is good, i.e., $\sigma_c\approx 0$, and we regress on $U_C$, then we have $(w_1,w_2)=(\frac{2\sigma^2_{d_2}}{2\sigma^2_{d_2}+\sigma^2}, \frac{\sigma^2}{2\sigma^2_{d_2}+\sigma^2})$, i.e., both $(w_1,w_2)$ will be the functions of domain-specific parameters, leading to failure to generalize to novel domains.
    \item If the channel condition is poor, i.e., $\sigma_c\gg 0$, no matter how many features we regress on, the coefficients $(w_1,w_2)$ will be the functions of channel condition parameters $\sigma_c$ and domain-specific parameters $(\sigma_{d_1},\sigma_{d_2})$, thus leading to failure to generalize to novel domains.
\end{itemize}

The results of only using $U_C$ yield the solution that is invariant to domain-specific parameters. In contrast, including $U_S$ in regression will make the coefficients be functions of these domain-specific parameters. Such correlation will result in the failure of generalizing to novel domains as the solutions of $\partial \mathcal{L}_{\text{mse}}/\partial \boldsymbol{w}=0$ will change with the domain-specific parameters, leading to a high MSE loss in the novel domain. 

According to the above example, to enhance generalization performance, task-oriented communication should concentrate on extracting and transmitting only causal features $U_C$, while avoiding transmission of information about spurious features $U_S$. Specifically, by only extracting and transmitting the causal features $U_C$, task-oriented communication can simultaneously reduce the communication overhead while also improving generalization performance. To achieve this goal, we should constrain the complexity of the extracted features, which is equivalent to constraining the mutual information between the extracted features and the input data, i.e., $I(\hat{Z},X)$. This aligns with the formulation of rate-distortion optimization in \eqref{eq:rd}. However, applying these principles becomes challenging with more complex tasks, where the underlying distribution of causal features, $p(U_C)$,  remains unknown. Consequently, rather than directly constraining the mutual information as outlined in \eqref{eq:rd}, we adopt the IB formulation in~\cite{tishby2000informationIB},

\begin{equation}\label{eq:IB}
    \min_{\theta,\phi} \mathcal{L}_{IB}=\underbrace{-I(\hZ,Y)}_{\text{Distortion}}+\beta \underbrace{I(X,\hZ)}_{\text{Rate}}.
\end{equation}
This strategy transforms the strict constraint into a more manageable Lagrange term, thus facilitating the optimization process.
The IB formulation represents a trade-off between rate and distortion, parameterized by $\beta$, where the first term seeks to maximize the mutual information between $\hZ$ and $Y$, i.e., minimizing the inference distortion. The second term, weighted by a positive $\beta$, quantifies how informative the received feature $\hZ$ is related to the original data $X$. Minimizing this term enables task-oriented communication system to distill the input data into a more compact and generalizable form.
\subsection{Invariant Principle Enhanced Information Bottleneck}
Simply constraining the complexity of the extracted features is insufficient for domain generalization, as spurious features may exhibit lower complexity compared to causal features. For instance, in the Colored-MNIST dataset, the color information (spurious feature) has lower entropy than the digit shape information (causal feature). If the task-oriented communication system is trained directly using the IB formulation specified in \eqref{eq:IB}, it may tend to utilize these spurious features as "shortcut" information for inference, thereby overlooking the causal features.
This observation motivates us to incorporate more effective constraints into the IB optimization process. As analyzed in Section~\ref{sec-model_problem}, the conditional mutual information terms satisfy $I(Y,D|U_C)=0$, whereas $I(Y,D|U_S) > 0$ and $I(Y,D|U_C, U_S) > 0$.
These conditions, as illustrated by our earlier examples, highlight the different influences of causal and spurious features. Specifically, given the causal features $U_C$, the only uncertainty to determine $Y$ comes from the annotation noise, i.e., $\sigma^2$ or $q$, \h{which does not change across different domains and yields $I(Y,D|U_C)=0$.}
\h{However, once the inferencer is given spurious feature $U_S$, the domain information $D$ will inevitably influence the inference result $Y$, which yields $I(Y,D|U_S) > 0$ and $I(Y,D|U_S,U_C)>0$.}


Building upon the principle introduced by~\cite{li2022invariant} and our previous analysis, we penalize the conditional mutual information $I(Y,D|\hat{Z})$ to ensure that only information from $U_C$ is transmitted. 
We minimize $I(Y,D|\hat{Z})$ instead of $I(Y,D|Z)$, because in both good or poor channel conditions, $I(Y,D|\hat{Z})\gtrsim I(Y,D|Z)$ generally holds. 
Given the constraint of minimizing $I(Y,D|\hat{Z})$, we define a new objective as follows,
\begin{equation}\label{eq:IFE}
    \min_{\theta,\phi} \underbrace{-I(Y,\hat{Z})}_{\text{Distortion}}+\beta \underbrace{I(X,\hat{Z})}_{\text{Rate}}+\lambda \underbrace{I(Y,D|\hat{Z}).}_{\text{Invariance}}
\end{equation}

To approximate the term $I(Y,D|\hat{Z})$, the dataset is partitioned into $D$ subsets, each denoted as $p(\x,\y|d),\, d\in\{1,...,D\}$ and $d$ is the domain index. Then $I(Y,D|\hat{Z})$ can be rewritten as

\begin{equation}
    \small
    \begin{aligned}
        I(Y,D|\hat{Z}) =& H(Y|\hZ) - H(Y|D,\hat{Z})\\ 
        =&\int \sum_{d}p(\x,\y,\z,d)[-\log(p_{\eta}(\y|\hz))\\ 
        &+\log(p_{\eta_d}(\y|\hz,d))]d\x d\y d\hz\\ 
        \overset{(a)}{=}&\int \sum_{d}p(d)p(\x,\y|d)p_{\theta}(\hz|\x)[-\log(p(\y|\hz))\\ 
        &+\log(p(\y|\hz,d))]d\x d\y d\hz\\ 
        \overset{(b)}{\approx}&\min_{\eta}\int p(\x,\y)p_\theta(\hz|\x)p_{\eta}(\hy|\hz)\mathcal{L}(\y,\hy)d\x d\y d\hz\\
        -\sum_d p(d)\min_{\eta_d}&\int p(\x,\y|d)p_{\theta}(\hz|\x)p_{\eta_d}(\hy|\hz,d)\mathcal{L}(\y,\hy)d\x d\y d\hz,
    \end{aligned}
\end{equation}
where $\mathcal{L}$ represents the cross entropy loss for classification tasks. Specifically, step (a) derives from the conditional independence relationship as shown in Fig. \ref{fig:pgm1}; 
and step (b) follows the minimization approach in~\cite{li2022invariant,farnia2016minimax}, transforming the conditional entropy into a classification loss with parameters $\eta$ and $\eta_d$.
Here, $\eta$ denotes the classification parameter for conditional distribution $p(\hy|\hz)$ over the whole dataset $p(\x,\y)$ and $\eta_d$ represents the classification parameter of conditional distribution $p(\hy|\hz,d)$ over $d$-th optimal dataset $p(\x,\y|d)$. 
However, such approximation will introduce additional parallel classifier $\eta_d$ which should be avoided in such resource-constrained scenario\footnote{The additional classifier either exists as a singular classifier, taking the domain index $d$ and received feature $\hz$ as input, or is multiple instances, determined by the quantity of domains.}. To address this issue, we further explore approximation techniques to simplify the objective.

Following~\cite{inference}, we assume that the optimal parameter $\eta_d$ for each domain $d$, characterized by the distribution $p(\x,\y|d)$, falls within the neighborhood of $\eta$, i.e., $\eta_d \in \{\eta_d|\Delta \eta_d =\eta-\eta_d,\Vert \Delta \eta_d \Vert^2_2\leq l\}$. Under this assumption, the conditional mutual information can be further approximated using Taylor expansion as follows,
\begin{equation}\label{eq:before_taylor}
    \begin{aligned}
        I(Y,D|\hat{Z})\approx& \min_{\eta} \mathbb{E}_{p(\x,\y)}[\mathcal{L}(\y,\hy|\eta,\theta)]\\
        &-\sum_{d}p(d)\min_{\eta_d:\Vert \Delta \eta_d\Vert^2_2\leq l} \mathbb{E}_{p(\x,\y|d)}[\mathcal{L}(\y,\hy|\eta_d,\theta)]\\
        =&\min_{\eta}\sum_{d}p(d)[\mathbb{E}_{p(\x,\y|d)}[\mathcal{L}(\y,\hy|\eta,\theta)]\\
        &-\min_{\Vert \Delta \eta_d \Vert^2_2\leq l}\mathbb{E}_{p(\x,\y|d)}[\mathcal{L}(\y,\hy|\eta+\Delta{\eta_d},\theta)]].\\
    \end{aligned}
\end{equation}
With the first-order Taylor approximation, we have

\begin{equation}
    \begin{aligned}
        \mathbb{E}[\mathcal{L}(\cdot|\eta+\Delta{\eta_d},\theta)] =& \mathbb{E}[\mathcal{L}(\cdot|\eta,\theta)]
        +(\nabla_{\eta}\mathbb{E}[\mathcal{L}(\cdot|\eta,\theta)])^T\Delta\eta_d\\&+o(\Delta_{\eta_d}).
    \end{aligned}
\end{equation}
Under the neighborhood constraint $\Vert \Delta \eta_d \Vert^2_2 \leq l$, the optimal $\Delta \eta_d$ that minimizes $\mathbb{E}_{p(\x,\y|d)}[\mathcal{L}(\y,\hy|\eta+\Delta{\eta_d},\theta)]$ is oriented along the direction of $-\nabla_{\eta}\mathbb{E}_{p(\x,\y|d)}[\mathcal{L}(\y,\hy|\eta,\theta)]$ with magnitude $l$. By substituting this into \eqref{eq:before_taylor}, we can obtain the gradient penalty term for the conditional mutual information,
\begin{equation}\label{eq:gradient_penalty}
    \begin{aligned}
         I(Y,D|\hat{Z}){\approx}&l\min_{\eta}\mathbb{E}_{d}[\Vert \nabla_{\eta} \mathbb{E}_{p(\x,\y|d)}[\mathcal{L}(\y,\hy|(\eta,\theta))]\Vert^2_2].\\
    \end{aligned}
\end{equation}
The transformation of an information-theoretical term into a gradient penalty allows the system to train just one classifier $\eta$ to estimate $I(Y,D|\hat{Z})$. Following the principle of joint estimation and optimization of mutual information in~\cite{hjelm2018learning}, the server-based network parameter $\phi$ can replace $\eta$ for estimation. Consequently, the final objective is as follows,
\begin{equation}\label{eq:IFE_gradient}
    \min_{\theta,\phi} \underbrace{-I(Y,\hat{Z})}_{\text{Distortion}}+\beta \underbrace{I(X,\hat{Z})}_{\text{Rate}}+\lambda \underbrace{\mathbb{E}_{p_d}[\Vert \nabla_{\phi} \mathbb{E}_{p(\x,\y|d)}[\mathcal{L}]\Vert^2_2]}_{\text{Invariance}}.
\end{equation}

\subsection{Variational Reformulation of Invariant Feature Encoding}
We now focus on computing the first two terms in \eqref{eq:IFE_gradient}, namely the rate-distortion terms, which can be formulated as follows,
\begin{equation}\label{eq:ib_expan}
    \begin{aligned}
            \mathcal{L}_{IB}=&-I(\hZ,Y)+\beta I(X,\hZ)\\
            =&\bE_{p(\x,\y)}[\bE_{p_{\theta}(\hz|\x)}[-\log p(\y|\hz)]\\
            &+\beta D_{KL}(p_{\theta}(\hz|\x)||p(\hz))]-H(Y).\\
    \end{aligned}
\end{equation}
The term $H(Y)$ can be omitted as a constant and the formulation includes two intractable high-dimensional integral with respect to distribution $p(\hz)$ and $p(\y|\hz)$,
\begin{equation}
    \begin{aligned}
        p(\hz) = & \int p(\boldsymbol{x})p_{\boldsymbol{\theta}}(\boldsymbol{\hat{z}}|\boldsymbol{x})d\boldsymbol{x}, \\
    p(\boldsymbol{y}|\boldsymbol{\hat{z}}) = & \int \frac{p(\boldsymbol{x},\boldsymbol{y})p_{\boldsymbol{\theta}}(\boldsymbol{\hat{z}}|\boldsymbol{x})}{p(\boldsymbol{\hat{z}})}d\boldsymbol{x}.
    \end{aligned}
\end{equation}

Following~\cite{vfe,alemi2016deepVIB}, we utilize two variational distributions $r(\hz)=N(0,\boldsymbol{I})$ and $q_{\phi}(\y|\hz)$ to approximate the true distribution $p(\hz)$ and $p(\y,\hz)$, respectively.
The parameter $\phi$ used here is consistent with that in the invariance term~\eqref{eq:IFE_gradient}. Based on these approximations, we define the VIFE objective as follows,

\begin{equation}\label{eq:vife}
    \begin{aligned}
        \min_{\theta,\phi}\mathcal{L}_{IFE} \leq &  \mathcal{L}_{VIFE}\\
        =&\mathbb{E}_{p_d}\big{\{}\mathbb{E}_{p(\x,\y|d)} \{\mathbb{E}_{p_{\theta}(\hz|\x)}[-\log q_{\phi}(\y|\hz)]\}\\
        &+\beta D_{KL}(p_{\theta}(\hz|\x)||r(\hz))\\
        &+\lambda \Vert \nabla_{\phi} \mathbb{E}_{p(\x,\y|d)}\{\mathbb{E}_{p_{\theta}(\hz|\x)}[-\log q_{\phi}(\y|\hz)]\}\Vert^2_2\big{\}}.
    \end{aligned}
\end{equation}
\h{The VIFE objective serves as an upper bound of the IFE objective for training the network and the proof is similar to~\cite{vfe}.}
\h{Finally, we implement the reparameterization trick~\cite{alemi2016deepVIB} to train the entire task-oriented communication system.}

\section{Label conditional feature Encoding for semantic-shift detection}\label{sec-ccib}
The VIFE framework developed in Section~\ref{sec-iib} addressed the generalization issues. Considering the existence of \textit{semantic-shift} data during deployment, in this section, we will extend our findings to semantic-shift scenario and propose a VLFE-based method to enhance the reliability of task-oriented communication. 

\subsection{Label-Conditional Feature Encoding}
The optimization of the IB and VIFE objectives uses a surrogate prior $r(\hz)$ for approximation. However, this surrogate prior may limit the system's ability to detect \textit{semantic-shifted} data, as all input data, including \textit{semantic-shifted} data, are mapped to the same latent distribution $r(\hz)$. \h{As a result}, \h{such a fixed prior issue renders the system struggle to distinguish whether the input data is \textit{semantic-shifted} or not.}

To address this limitation, we propose an LFE-based scheme, which incorporates a label-dependent prior into the objective. This is achieved by modifying the rate-distortion term in the optimization of IB or IFE to a conditional mutual information formulation,
\begin{equation}\label{eq:lfe}
\mathcal{L}_{LFE}= - \underbrace{I(Y;\hat{Z})}_{\text{Distortion}}+\underbrace{\beta I(X;\hat{Z}|Y)}_{\text{Rate}}.
\end{equation}

\begin{figure}[t]  
  \centering         
  \subfloat[Objective of IB in~\eqref{eq:IB}]{\includegraphics[width=0.45\linewidth]{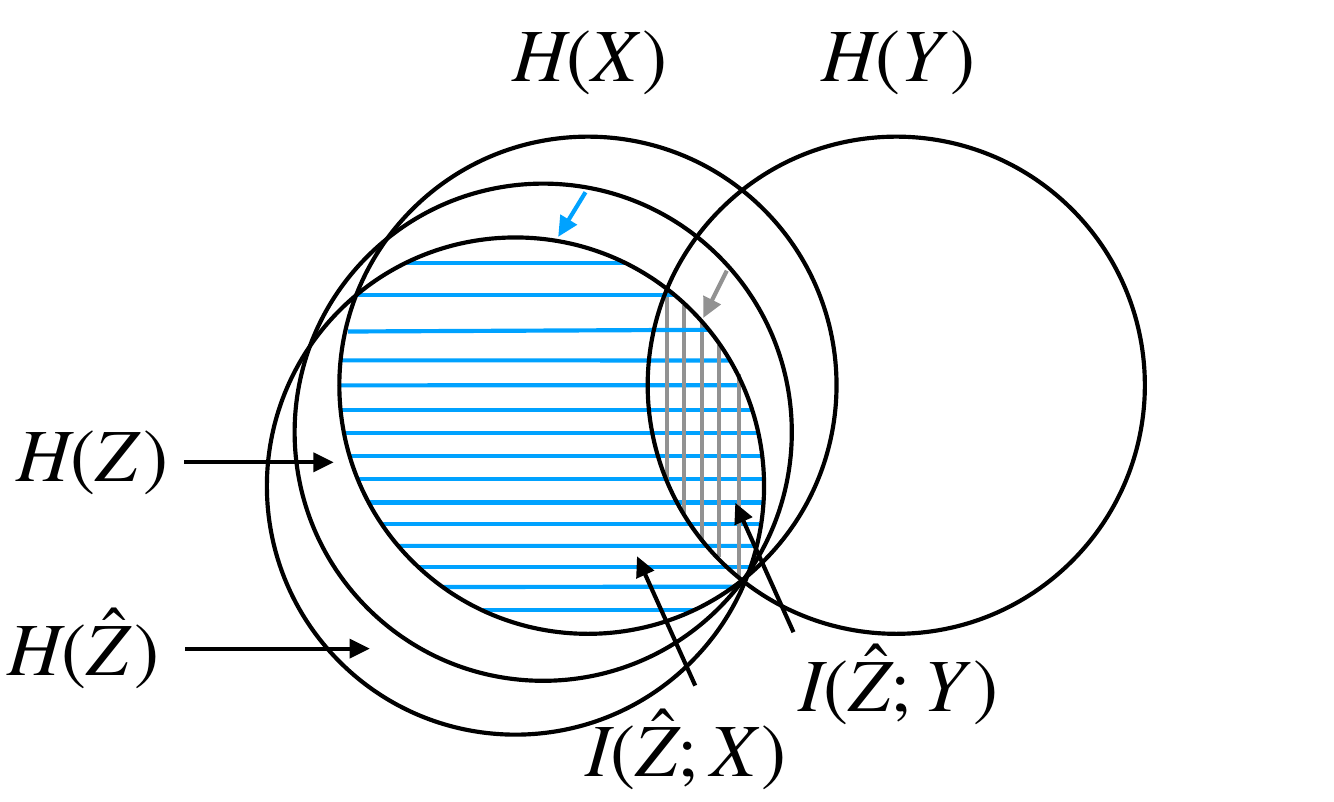}}\label{fig:ib_venn}
  \subfloat[Objective of LFE in~\eqref{eq:lfe}]{\includegraphics[width=0.45\linewidth]{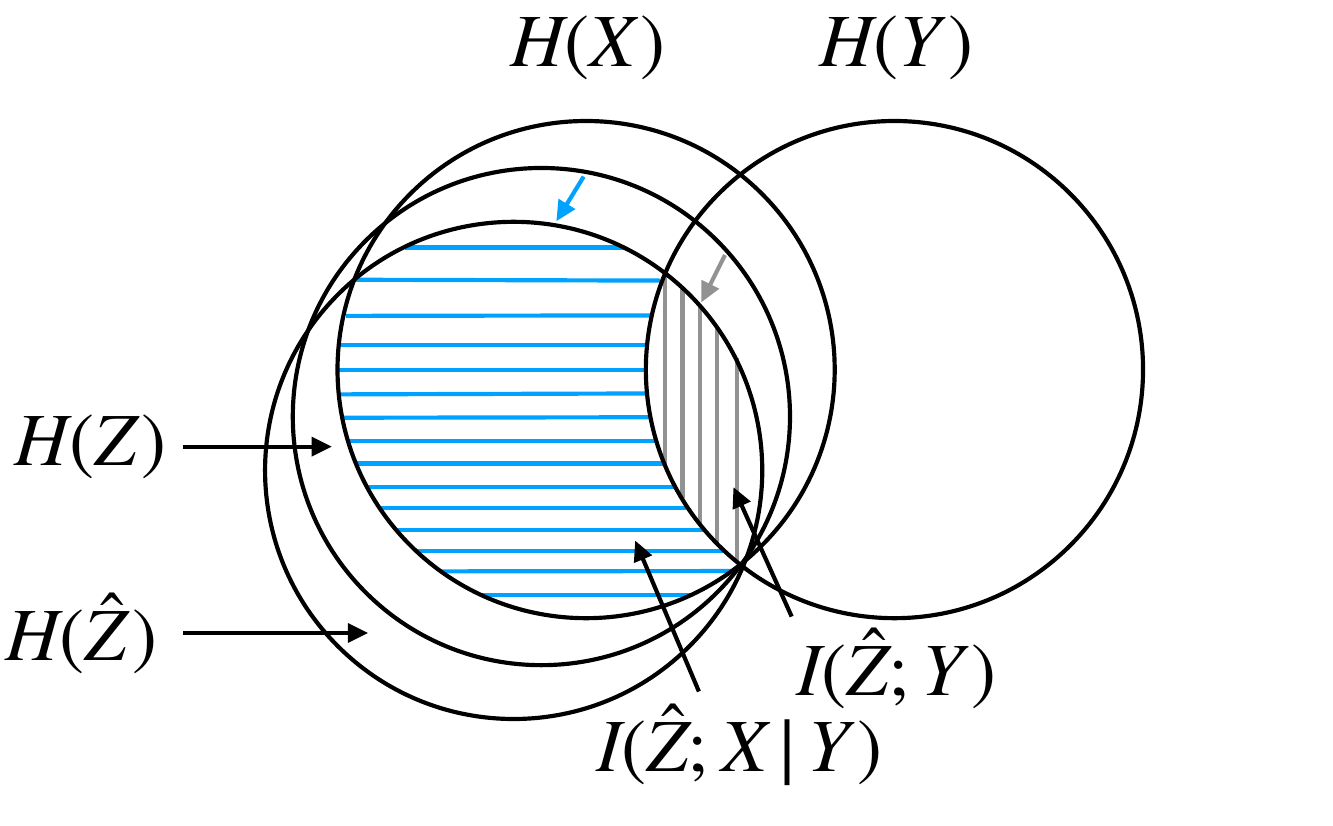}}\label{fig:ccib_venn}
  \caption{The Venn diagrams of IB and CCIB show optimization (maximum and minimum) areas in their objectives. Blue lines: the area being minimized by both objectives. Gray lines: the area being maximized by both objectives. The colored arrows indicate the information loss because of the channel noise. \textbf{IB}: The term to be maximized, i.e., $I(\hat{Z};Y)$, is a subset of the term to be minimized, namely, $I(\hat{Z};X)$. \textbf{LFE}: The term to be maximized, i.e., $I(\hat{Z};Y)$, is no longer a subset of the term to be minimized, namely, $I(\hat{Z};X|Y)$.}
  \label{fig:ib_ccib_venn}         
\end{figure}

This formulation offers two significant benefits. First, by minimizing the conditional mutual information $I(X;\hat{Z}|Y)$, it will not compromise the maximization of $I(Y;\hat{Z})$. This is shown in the Venn diagrams in Fig.\,\ref{fig:ib_ccib_venn}, where the areas to be minimized and maximized by IB and LFE are illustrated, respectively, indicating that LFE avoids the conflict between minimizing $I(X;\hat{Z})$ and maximizing $I(Y;\hat{Z})$.
Second, the conditional approach facilitates better \textit{semantic-shift} detection. This is achieved by designing the latent space to distinctly separate features based on their labels, which aligns with the strategies in the existing literature to enhance semantic-shift detection~\cite{yang2021generalized}. Such label-dependent priors allow for efficient differentiation of ID data based on their labels and the \textit{semantic-shifted} data are likely to fall into areas of the latent space that are not covered by the label-dependent priors, thereby making them easier to detect.

We further define the VLFE objective as an upper bound of $\mathcal{L}_{LFE}$ with
\begin{equation}
    \begin{aligned}
        \mathcal{L}_{LFE}\leq & \mathcal{L}_{VLFE} \\
        =&\mathbb{E}_{p(\x,\y)}\{\mathbb{E}_{p_{\theta}(\hz|\x)}[-\log q_{\phi}(\y|\hz)]\\
        &+\beta D_{KL}(p_{\theta}(\hz|\x)||r(\hz|\y))\},
    \end{aligned}
    \label{eq:vlfe}
\end{equation}
where $r(\hz|\y)$ and $q_{\phi}(\y|\hz)$ are variational approximation of the true posterior distribution $p(\y|\hz)$ and true conditional latent distribution $p(\hz|\y)$, respectively. The proof of the VLFE objective as an upper bound of the LFE objective can be found in Appendix~\ref{app:vlfe}.

\h{\subsection{Contrastive Learning-Based Latent Space Separation and Combination with VIFE}}
To effectively utilize conditional priors for separating the latent space of various ID data, we develop a contrastive learning-based approach with label-dependent variational priors. The latent space of each class is modeled as a Gaussian distribution, $r(\hz|\y=\boldsymbol{c})=N(\hz|\mu_c,\Sigma_c)$ for class $\boldsymbol{c}$. Different from the approaches that utilize predefined conditional priors~\cite{sun2020conditional} and inspired by~\cite{vmdls}, our method dynamically assigns these priors based on the latent features observed in the previous epoch. Specifically, we compute the label-conditional mean and covariance from the $(t-1)$-th epoch as,
\begin{equation}
    \begin{aligned}
        \boldsymbol{\mu}'_{c,t-1}&=\mathbb{E}[\mathbf{\hat{Z}}|\y=\boldsymbol{c}]\\
    \boldsymbol{\Sigma}'_{c,t-1}&=\mathbb{E}[\mathbf{\hat{Z}\hat{Z}^T}|\y=\boldsymbol{c}]-\boldsymbol{\mu}'_{c,t-1}\boldsymbol{\mu}'^T_{c,t-1}.
    \end{aligned}
\end{equation}
For the subsequent $t$-th epoch, the conditional latent prior for each class is updated as 

\begin{equation}
    r_t(\mathbf{\hat{z}}|\y=\boldsymbol{c})=\mathcal{N}(\mathbf{\hat{z}}|\boldsymbol{\mu}'_{c,t-1},\boldsymbol{\Sigma}'_{c,t-1}).
\end{equation}

To further enforce the separation among the latent spaces of different classes, we employ a triplet loss, which is widely used in contrastive learning to enhance the distinction between data representations of various labels~\cite{schroff2015facenet}. The triplet loss $\mathcal{L}_t$ is defined as,
\begin{equation}\label{eq:triplet loss}
\begin{aligned}
    \mathcal{L}_t =& \frac{1}{T}\frac{1}{C}\sum_{c=1}^{C}\sum_{t=1}^{T}\max(||\mathbf{\hat{z}}_{r,c,t}
    -\mathbf{\hat{z}}_{m,c,t}||^2_2\\
    &-||\mathbf{\hat{z}}_{r,c,t}-\mathbf{\hat{z}}_{n,-c,t}||^2_2+\alpha,0),
\end{aligned}
\end{equation}
where $T$ is the total number of triads in the dataset, $C$ is the number of known classes, and $\alpha$ is a predefined margin. The triplet loss mechanism strives to decrease the distance between a reference latent representation $\mathbf{\hat{z}}_{r,c}$ and a matching representation $\mathbf{\hat{z}}_{m,c}$ from the same class $c$, while increasing the distance between the reference and a non-matching representation $\mathbf{\hat{z}}_{n,-c}$ from a different class. Thus the triplet loss can encourage the latent space to be separated by class labels, which is crucial for detecting \textit{semantic-shifted} data~\cite{vmdls}.

\h{Finally, we have the contrastive learning-based VLFE objective given by} 
\begin{algorithm}[t]
    \small
    \caption{Training Algorithm for VIFE-VLFE}
    \begin{algorithmic}[1]
    \Require $T$ (number of epochs), $D$ (number of data domain), $L$ (number of channel noise samples per datapoint), batch size $B$ and channel variance $\sigma^{2}$.
    \While{epoch $t=1$ to $T$}
    \While{domain $d=1$ to $D$}
    \State Select batch data $\left\{\left(\boldsymbol{x_{i}}, \boldsymbol{y_{i}}\right)|d\right\}_{i=1}^{B}$ from $d$-th domain
    \State Sample the channel noise $\{\sigma^2_l\}_{l=1}^L\sim\mathcal{N}(0,\sigma^2 \boldsymbol{I})$
    \State Compute the KL-divergence and contrastive loss based 
    \Statex {\qquad \quad on~\eqref{eq:triplet loss}}
    \State Compute gradient penalty based on~\eqref{eq:gradient_penalty}
    \EndWhile
    \State Compute $\mathcal{L}_{\text{final}}$ over all domain data based on (\ref{eq:final})
    \State Update parameters $\boldsymbol{\phi},\boldsymbol{\theta}$ through backpropagation.
    \EndWhile
    \end{algorithmic}
    \label{alg:ife_lfe}
\end{algorithm}
\begin{align}\label{eq:mc_vlfe}
        &\mathcal{L}_{VLFE} = \frac{1}{N}\sum_{n=1}^N(\beta D_{KL}(p_{\theta}(\mathbf{\hat{z}}|\mathbf{x}_n)||r(\mathbf{\hat{z}}|y_n))\notag \\
    &-\frac{1}{L}\sum_{l=1}^{L} [\log q_{\phi}(y_n|\mathbf{\hat{z}})])+\frac{1}{T}\frac{1}{C}\sum_{c=1}^{C}\sum_{t=1}^{T}\max(||\mathbf{\hat{z}}_{r,c,t}\notag \\
    &-\mathbf{\hat{z}}_{m,c,t}||^2_2-||\mathbf{\hat{z}}_{r,c,t}-\mathbf{\hat{z}}_{n,-c,t}||^2_2+\alpha,0),
\end{align}
where $N$ denotes the total number of data points and $L$ represents the number of channel noise samples per data point.

During the testing phase, the final conditional priors $\{\boldsymbol{\mu}_c,\boldsymbol{\Sigma}_c\}^C_{c=1}$ are stored for \textit{semantic-shifted} data detection. Given a data point $\mathbf{x}$, the system calculates the log-likelihood score of the latent features $\hz$ with respect to all known classes, utilizing the stored Gaussian distributions. These scores are further weighted by test-point reconstruction techniques~\cite{sun2020conditional}. If the maximum log-likelihood score falls below a predefined threshold $\tau$, the system identifies the data point as \textit{semantic-shifted}. This threshold-based approach allows the system to effectively discern significant deviations in data characteristics that may indicate semantic shifts, thereby enabling robust detection of out-of-distribution data.

To simultaneously generalize to \textit{domain-shifted} data and detect \textit{semantic-shited} data, we combine the VIFE and VLFE objectives to jointly train the task-oriented communication system. This is achieved by adding the gradient penalty to the objective of $\mathcal{L}_{VLFE}$ to obtain
\begin{equation}
    \begin{aligned}
        \mathcal{L}_{\text{final}} = \mathcal{L}_{VLFE}+\lambda \Vert \mathbb{E}_{p_d}[\Vert \nabla_{\phi} \mathbb{E}_{p(\x,\y|d)}[\mathcal{L}]\Vert^2_2] \Vert^2_2,
    \end{aligned}
    \label{eq:final}
\end{equation}
and the training procedure is outlined in Algorithm~\ref{alg:ife_lfe}.

\vspace{0.3cm}
\section{Experimental Results}\label{sec-exp}
In this section, we evaluate the proposed task-oriented communication scheme on edge-assisted remote image classification tasks to verify the effectiveness of the proposed method on \textit{domain-shift} generalization and \textit{semantic-shift} detection. Furthermore, the ablation studies verify the robustness of hyperparameters selection for the proposed method\footnote{The code will be available at {github.com/hlidmhkust}.}.

\subsection{Simulation Setup}
\subsubsection{Datasets and Metrics}
To verify the performance on domain-shifted data, we evaluate the proposed method on two vision classification datasets with spurious features, i.e., Colored-MNIST~\cite{arjovsky2019invariant} and Colored-Object~\cite{ahmed2020systematic}. The correlation between spurious features and causal features is depicted by the \textit{spurious bias ratio} $s$, \g{i.e., each digit or object has an associated color with probability $s$, while assigning equal probability to other colors with a combined probability of $1-s$.}
We use two bias ratios for training and one bias ratio for testing, denoted as $(s_{tr_1},s_{tr_2},s_{te})$.
For the Colored-MNIST dataset, the bias ratios are set to $(0.9,0.8,0.1)$ and the label noise is considered, i.e., the label is randomly changed with probability $\rho=25\%$. For the Colored-Object dataset, the bias ratios are set to $(0.999,0.7,0.1)$ and the label noise is $\rho=10\%$. Furthermore, the generalization performance metric is top-1 accuracy on the testing dataset, and a higher accuracy indicates a better performance in classification. Following the same settings in~\cite{vfe}, we set the symbol rate as 9,600 Baud.

To verify the performance of the proposed method in detecting semantic-shifted data, we evaluate the proposed method on LSUN dataset~\cite{wang2017knowledge} and ImageNet dataset~\cite{deng2009imagenet}, where models are trained on Colored-MNIST and Colored-Object dataset, respectively. Moreover, the performance metric for detection is the area under the receiver operating characteristic curve (AUROC), and a higher AUROC value indicates a superior detection performance.

\subsubsection{Baselines}
\begin{figure*}[t]
    \centering
    \captionsetup[subfigure]{labelformat=empty}
    \subfloat[(a) Colored-MNIST, $\textrm{PSNR}= 10$ dB]{
        \centering
        \includegraphics[width=0.46\linewidth]{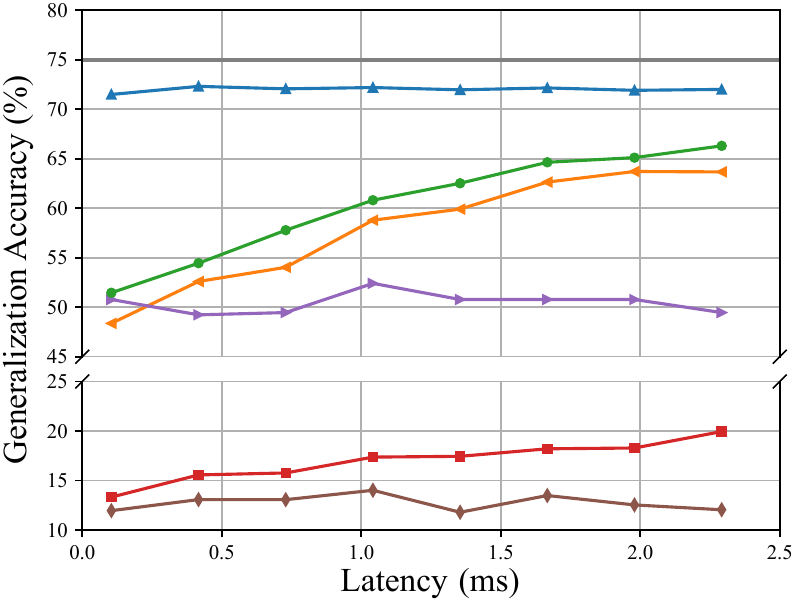}
    }
    \hfill
    \subfloat[(b) Colored-MNIST, $\textrm{PSNR}= 20$ dB]{
        \centering
        \includegraphics[width=0.46\linewidth]{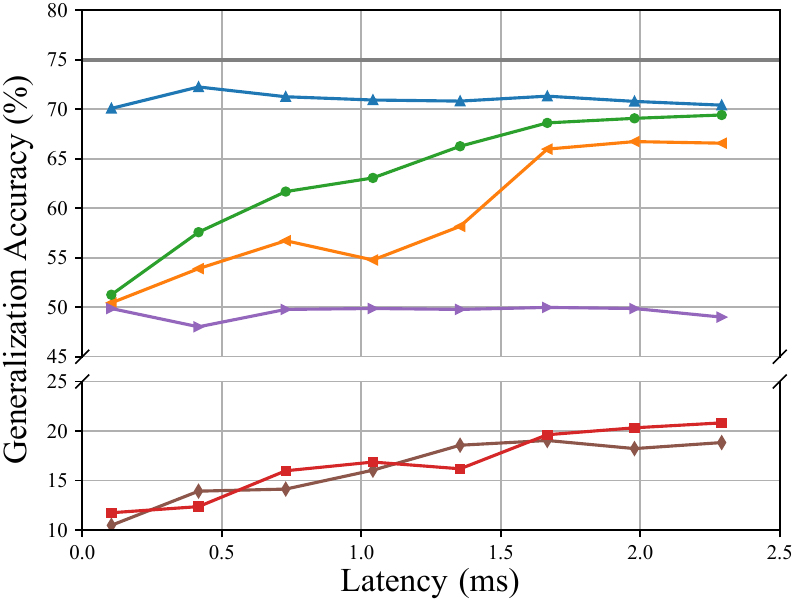}
    }
    
    \subfloat[(c) Colored-Object, $\textrm{PSNR}= 10$ dB]{
        \centering
        \includegraphics[width=0.46\linewidth]{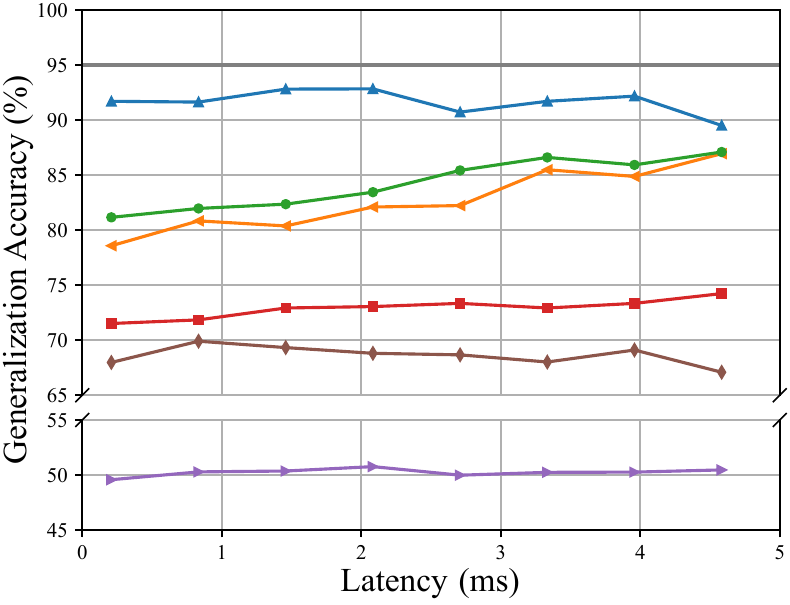}
    }
    \hfill
    \subfloat[(d)Colored-Object, $\textrm{PSNR}= 20$ dB]{
        \centering
        \includegraphics[width=0.46\linewidth]{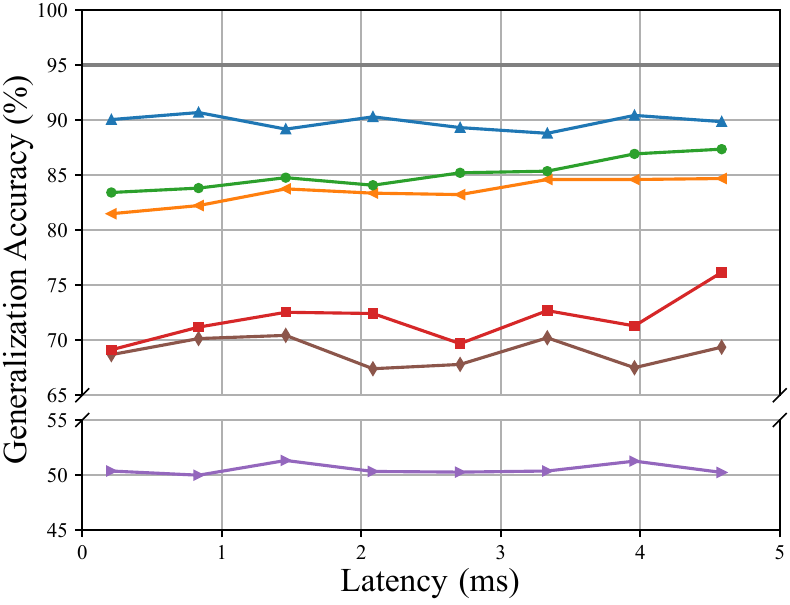}
    }
    
    \subfloat[]{
        \centering
        \includegraphics[width=0.95\linewidth]{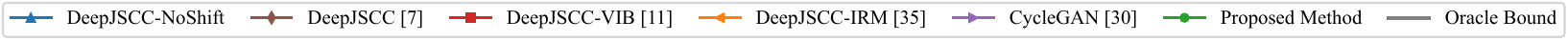}
    }
    \caption{The rate-distortion curves for image classification tasks with $\text{PSNR}_{\text{test}}=\text{PSNR}_{\text{train}}$. (a) Colored-MNIST, $\textrm{PSNR}= 10\textrm{dB}$, (b) Colored-MNIST, $\textrm{PSNR}= 20\textrm{dB}$, (c) Colored-Object, $\textrm{PSNR}= 10\textrm{dB}$ and (d) Colored-Object, $\textrm{PSNR}= 20\textrm{dB}$.
    }
    \label{fig:rd_curve}
\end{figure*}

\begin{figure*}[t]
    \centering
    \captionsetup[subfigure]{labelformat=empty}
    \subfloat[(a) Colored-MNIST, $\textrm{PSNR}_{\text{train}}= 10$ dB]{
        \centering
        \includegraphics[width=0.44\linewidth]{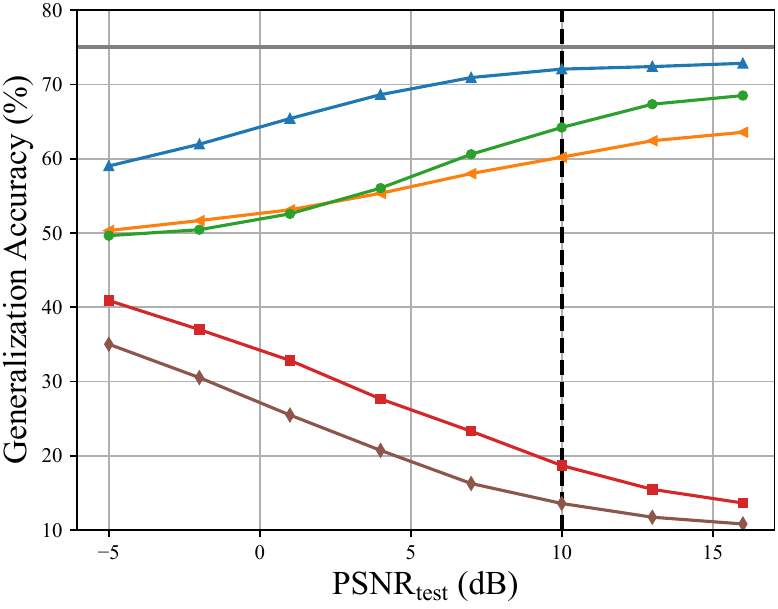}
    }
    \hfill
    \subfloat[(b) Colored-MNIST, $\textrm{PSNR}_{\text{train}}= 20$ dB]{
        \centering
        \includegraphics[width=0.44\linewidth]{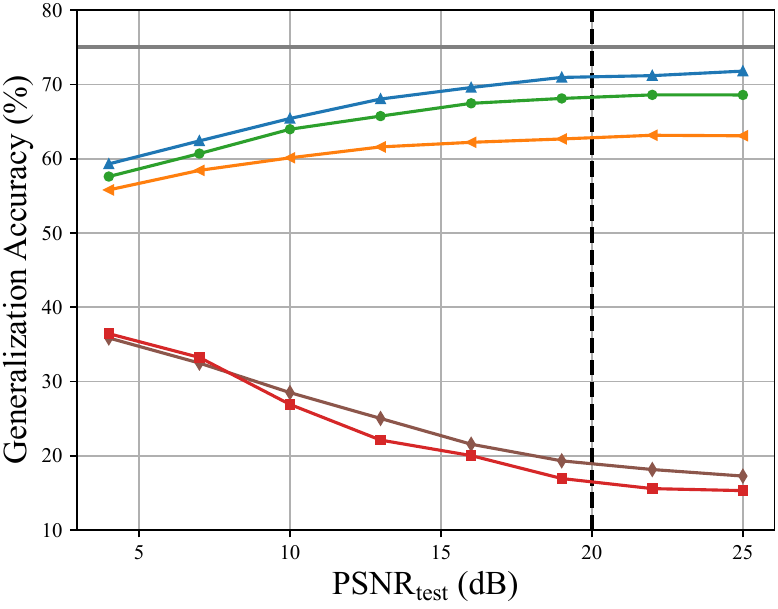}
    }

    \subfloat[(c) Colored-Object, $\textrm{PSNR}_{\text{train}}= 10$ dB]{
        \centering
        \includegraphics[width=0.44\linewidth]{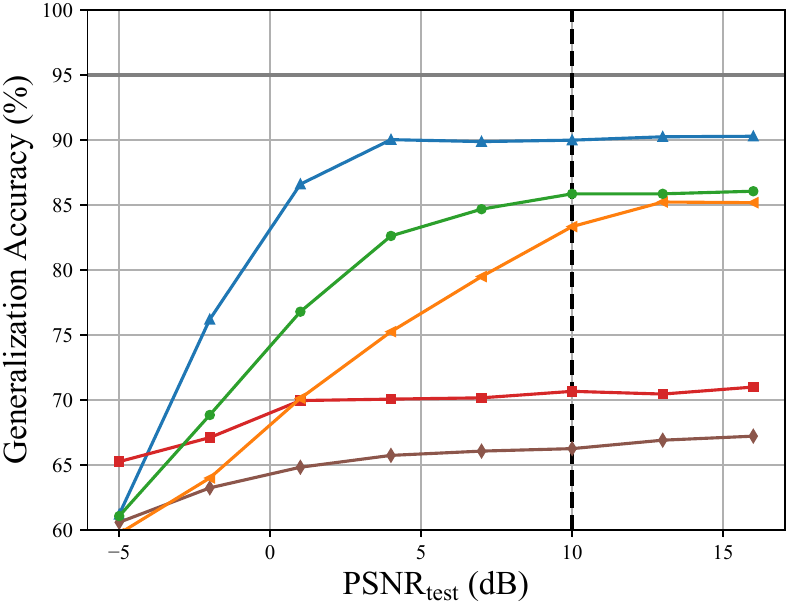}
    }
    \hfill
    \subfloat[(d)Colored-Object, $\textrm{PSNR}_{\text{train}}= 20$ dB]{
        \centering
        \includegraphics[width=0.44\linewidth]{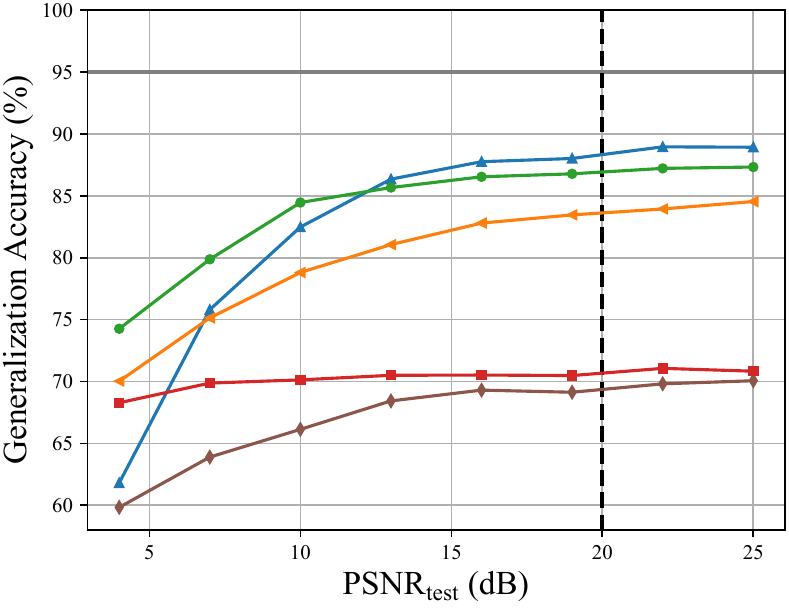}
    }

    \subfloat[]{
        \centering
        \includegraphics[width=0.95\linewidth]{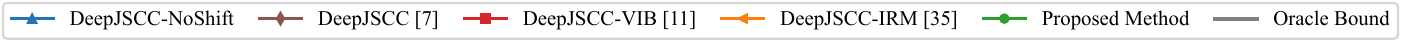}
    }
    \caption{Performance of image classification tasks over different $\textrm{PSNR}_{\text{test}}$. All methods are trained on a specific $\textrm{PSNR}_{\text{train}}$. (a) Colored-MNIST, $\textrm{PSNR}_{\text{train}}= 10\textrm{dB}$, (b) Colored-MNIST, $\textrm{PSNR}_{\text{train}}= 20\textrm{dB}$, (c) Colored-Object, $\textrm{PSNR}_{\text{train}}= 10\textrm{dB}$ and (d) Colored-Object, $\textrm{PSNR}_{\text{train}}= 20\textrm{dB}$.
    }
    \label{fig:rd_curve_robust}
\end{figure*}
For performance comparison in terms of generalization, the following baselines are compared:
\begin{itemize}
    \item \textbf{DeepJSCC}: The vanilla learning-based strategy for data transmission and task inference~\cite{jssc_deniz,jankowski2020wireless}.
    \item \g{\textbf{DeepJSCC-NoShift}: DeepJSCC-NoShift is trained with the same training and testing data distribution without any shifts.}
    \item \textbf{DeepJSCC-IRM}: DeepJSCC-IRM~\cite{arjovsky2019invariant} is directly trained with IRM principle which aims to find the causal relation between the input and target data to achieve generalization purpose.
    \item \textbf{DeepJSCC-VIB}: DeepJSCC-VIB is proposed in~\cite{vfe} which utilizes the variational IB (VIB) to learn the task-relevant information. Our method is the same as DeepJSCC-VIB when excluding IRM method.
    \item \g{\textbf{CycleGAN}: This method is proposed in~\cite{zhang2022deep} which requires \textit{retraining} on the testing dataset to convert the testing input into a similar training sample with CycleGAN.}
    \item \textbf{Oracle bound}: Oracle bound is the classifier with perfect knowledge about causal relationship between the input and target data and can achieve $1-\rho$ accuracy with label noise $\rho$.
\end{itemize}

\begin{table}[t]

    \selectfont
    \caption{The generalization accuracy under different training PSNR with fixed transmission latency for Colored-MNIST and Colored-Object classification tasks.}
    \begin{center}
    \scalebox{0.95}{
    \begin{tabular}{@{}c|cccccc@{}}
    \toprule \textbf{Colored-MNIST} & 4 dB & 7 dB & 10 dB & 13 dB & 16 dB & 19 dB \\
    \midrule
    DeepJSCC-NoShift & 71.56 & 71.98 & 72.09 & 71.96 & 71.41 & 70.89  \\
    DeepJSCC & 10.93 & 11.16 & 13.44 & 15.66 & 18.05 & 19.85  \\
    DeepJSCC-VIB & 14.44 & 13.36 & 18.20 & 18.00 & 16.78 & 14.78  \\
    DeepJSCC-IRM & 54.90 & 59.34 & 62.64 & 63.05 & 65.61 & 66.88  \\
    \textbf{Proposed} & \textbf{56.87} & \textbf{61.22} & \textbf{64.64} & \textbf{66.50} & \textbf{67.71} & \textbf{68.39}  \\
    \midrule\textbf{Colored-Object} & 4 dB & 7 dB & 10 dB & 13 dB & 16 dB & 19 dB\\
    \midrule DeepJSCC-NoShift & 91.50 & 90.65 & 90.22 & 89.04 & 91.22 & 90.43 \\
    DeepJSCC & 64.80 & 70.19 & 71.89 & 70.76 & 65.85 & 71.80 \\
    DeepJSCC-VIB & 70.81 & 73.85 & 75.37 & 71.76 & 73.87 & 72.98 \\
    DeepJSCC-IRM & 85.70 & 84.28 & 83.19 & 86.19 & 84.57 & 85.20  \\
    \textbf{Proposed} & \textbf{85.72} & \textbf{87.17} & \textbf{87.37} & \textbf{86.54} & \textbf{86.59} & \textbf{85.44} \\
        \bottomrule
    \end{tabular}}
    
    \label{table-awgn}
    \end{center}
    \end{table}

For detection performance comparison, the following baselines are considered:

\begin{itemize}
    \item \textbf{DeepJSCC-ODIN}: \g{This method uses the feature extracted by DeepJSCC scheme and ODIN detection technique~\cite{hsu2020generalized} to detect the semantic-shifted data.}
    \item \textbf{VIB-ODIN}:
    \g{This method uses the feature extracted by VIB scheme~\cite{vfe} and ODIN detection technique~\cite{hsu2020generalized} to detect the semantic-shifted data.}
\end{itemize}

\subsection{Generalization Performance}
In this subsection, we investigate the generalization performance of the proposed method in static channel conditions where the channel has the same PSNR for training and testing, denoted as $\text{PSNR}_{\text{train}}=\text{PSNR}_{\text{test}}\in\{10\ \text{dB}, 20\ \text{dB}\}$. We show the generalization accuracy, which is evaluated only on test domain dataset with bias ratio $s_{te}$ and different transmission latency. We change the hyperparameters $\lambda\in[10^3,10^6]$ and $\beta\in[10^{-5},10^{-2}]$ for both datasets. We follow the test-domain validation model selection protocol~\cite{gulrajani2020search} to select the best hyperparameters. The transmission latency is determined by the dimension of transmitter output. The rate-distortion tradeoff curves of all methods are shown in Fig.\,\ref{fig:rd_curve} with Colored-MNIST and Colored-Object classification tasks, respectively.

\g{From Fig.\,\ref{fig:rd_curve}, we observe that the non-causal-aware methods, i.e., DeepJSCC and DeepJSCC-VIB, achieve similarly worst performance in both datasets. This is because the task-irrelevant information containing misleading spurious features, i.e., the digit color in Colored-MNIST and background color in Colored-Object, are unreliable for predicting the target label. When the correlation between spurious features and the label changes, the generalization performance of DeepJSCC and DeepJSCC-VIB will be significantly degraded. 
Only when there is no distribution shift in the test data, as illustrated by the results of DeepJSCC-NoShift, can the generalization performance be guaranteed to approach the oracle bound. In contrast, the proposed method and DeepJSCC-IRM achieve an excellent performance even with distribution shifts. This is because the proposed method and DeepJSCC-IRM can effectively learn the task-relevant information and remove the task-irrelevant information. 
Furthermore, we observe that the proposed method achieves the best rate-distortion tradeoff among all methods. This indicates the proposed method can effectively learn the task-relevant information and reduce unnecessary information transmission for task-inference\footnote{The convergence of CycleGAN-based method is dramatically unstable, as also discussed in ~\cite{zhu2017unpaired, mescheder2018training}. We will not include this method in the subsequent simulations.}.} 

Next, we evaluate the generalization performance of all methods with different PSNR. Specifically, we control the transmission latency for Colored-MNIST is under 1 ms and that for Colored-Object is under 8 ms, respectively. 

\g{Table\,\ref{table-awgn} shows the generalization accuracy of baselines and the proposed method under different channel conditions. It is observed that the proposed method consistently outperforms all baselines in all PSNR settings, implying that the proposed method can effectively identify task-relevant information and achieve resilient generalization performance under different channel conditions.}

\subsection{Robustness to Dynamic Channel Conditions}

\begin{figure}[ht]
    \centering
    \includegraphics[width=0.95\linewidth]{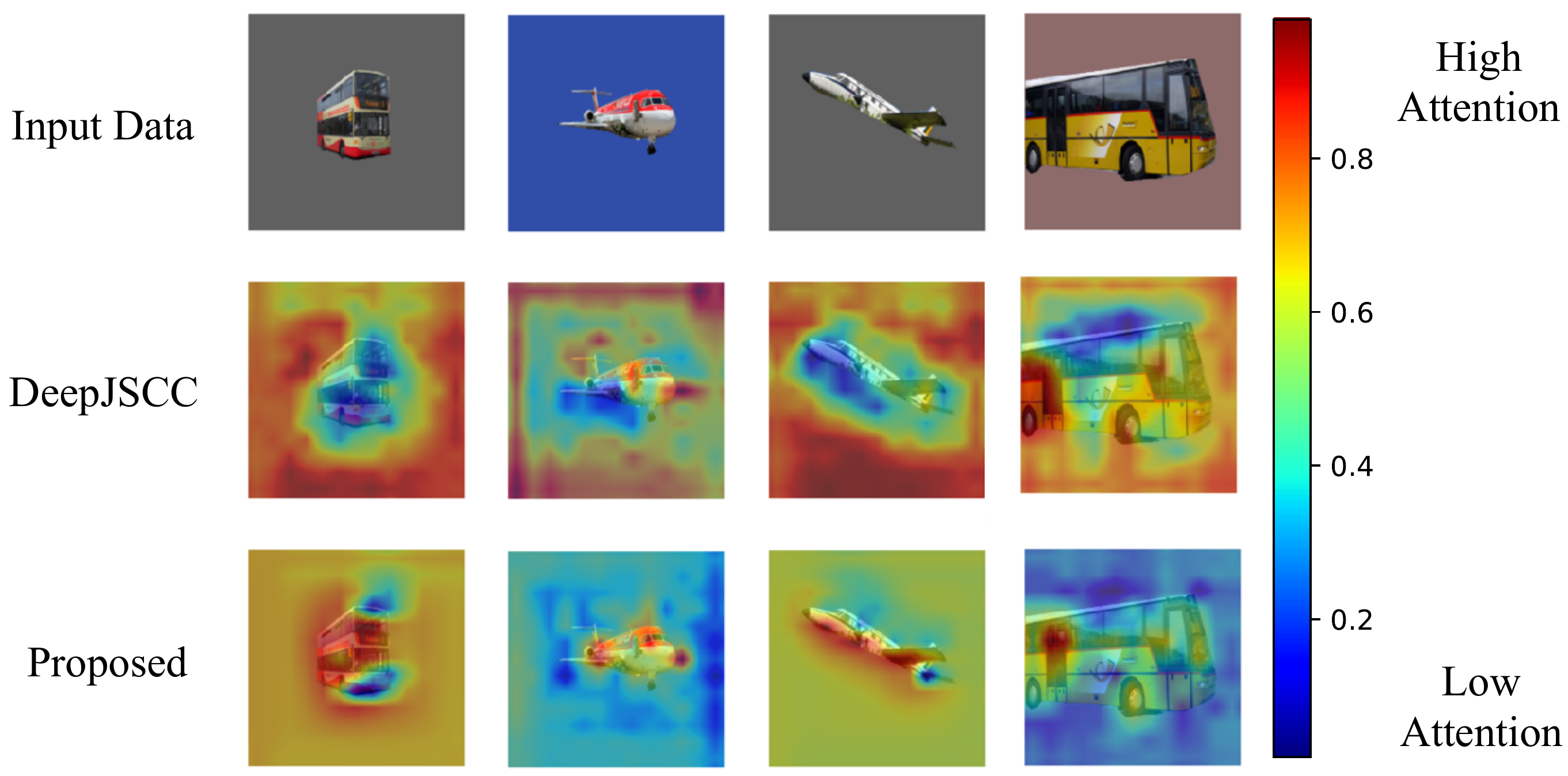}
    \caption{The attention heatmap of DeepJSCC and proposed method.}
    \label{fig:att_map}
\end{figure}
In this subsection, we evaluate the robustness of the proposed method in dynamic channel conditions. All the experiments are trained in an end-to-end manner with specific $\text{PSNR}_{\text{train}}\in \{10\ \text{dB},20\ \text{dB}\}$. We further test the generalization performance of all evaluated methods under $\text{PSNR}_{\text{test}}\in [-5\ \text{dB},25\ \text{dB}]$. 

\g{Fig.\,\ref{fig:rd_curve_robust} illustrates the generalization performance of all methods under different $\text{PSNR}_{\text{test}}$. It is shown that the proposed method consistently outperforms baselines, indicating the proposed method is more robust than all baselines in dynamic channel conditions.
Interestingly, the generalization performance of non-causal-based methods, DeepJSCC and DeepJSCC-VIB, degrades as the PSNR improves on the Colored-MNIST dataset. This degradation is attributed to the transmission of misleading task-irrelevant information by these methods. Consequently, the more accurate the information is recovered at the receiver, the worse the system performance will be.
Furthermore, we observe that under low $\text{PSNR}_{\text{test}}$, the generalization performance of DeepJSCC-NoShift is worse than DeepJSCC-IRM, DeepJSCC-VIB and the proposed method in Colored-Object classification tasks. This is because IRM-based and IB-based methods also have generalizability to channel noise distribution shifts, thus can combat the channel noise.}

\begin{figure*}[t]
    \centering
    \subfloat[Colored-MNIST with different $\beta$]{
        \centering
        \includegraphics[width=0.47\linewidth]{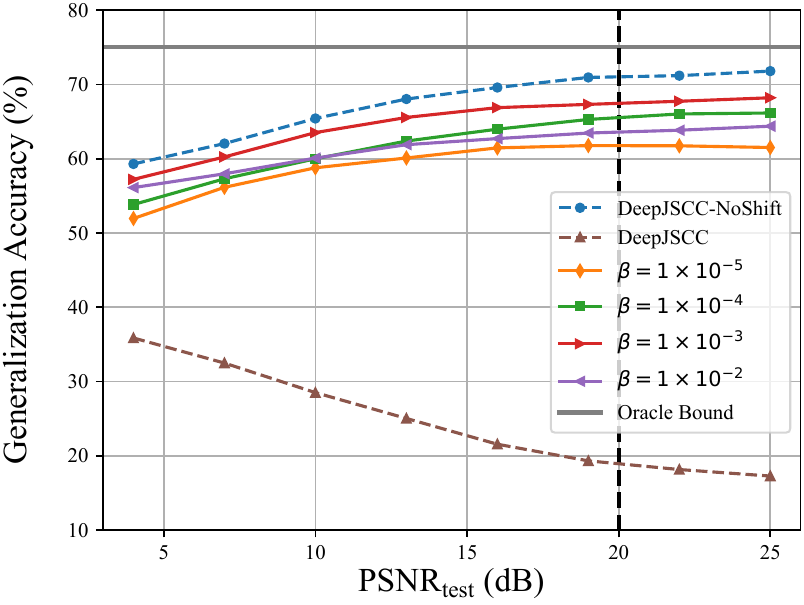}
    }
    \hfill
    \subfloat[Colored-Object with different $\beta$]{
        \centering
        \includegraphics[width=0.47\linewidth]{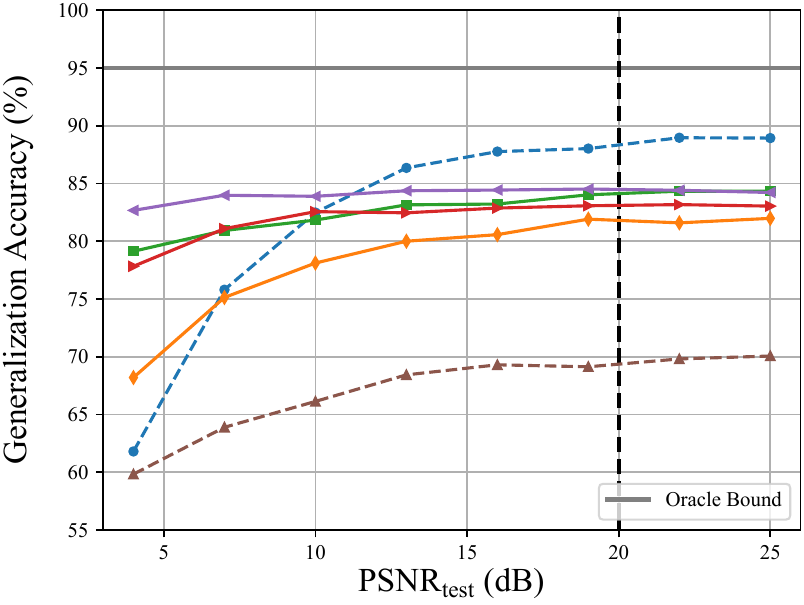}
    }
    \vfill
    \subfloat[Colored-MNIST with different $\lambda$]{
        \centering
        \includegraphics[width=0.47\linewidth]{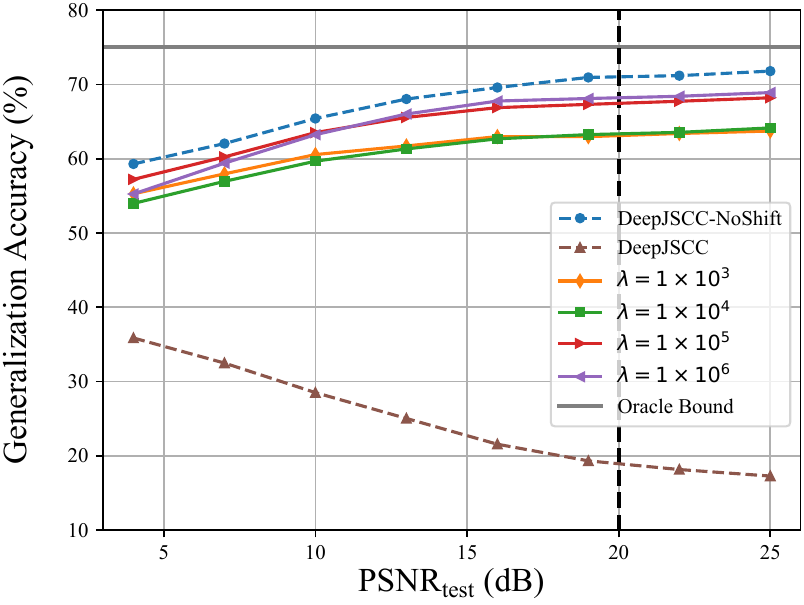}
    }
    \hfill
    \subfloat[Colored-Object with different $\lambda$]{
        \centering
        \includegraphics[width=0.47\linewidth]{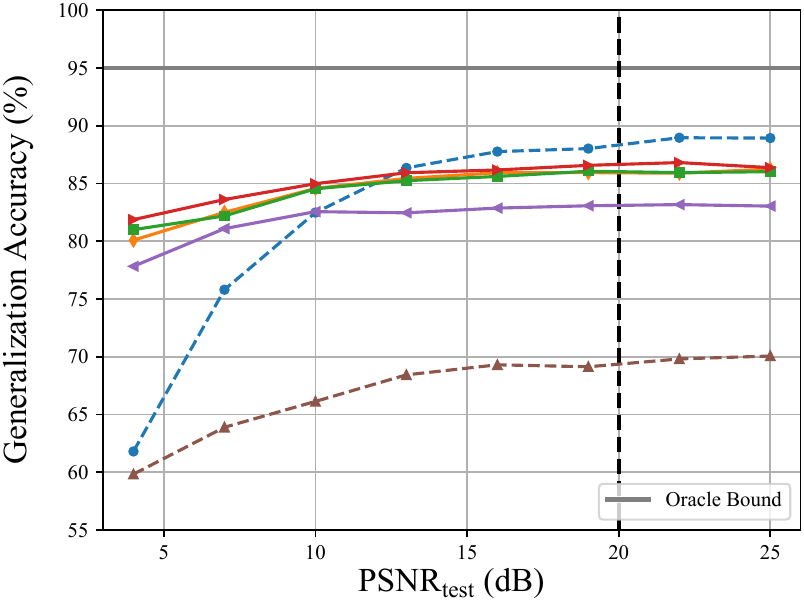}
    }
    \caption{Ablation study for hyperparameter robustness of the proposed method. (a), (b) performance of different $\beta$ on Colored-MNIST and Colored-Object, (c), (d) performance of different $\lambda$ on Colored-MNIST and Colored-Object.
    }
    \label{fig:ablation}
\end{figure*}

\subsection{Attention Heatmap Visualization}

\g{In Fig.\,\ref{fig:att_map}, we visualize the attention heatmap for the DeepJSCC and proposed method on Colored-Object dataset. The attention heatmap can be obtained by class activation mapping (CAM) method~\cite{zhou2016learning}. It can visualize the attention of the model on the input image, which illustrates the information that the model focuses, encodes and transmits. 
It is shown that the proposed method can focus on the task-relevant information, i.e., the object, while the DeepJSCC method focuses on the task-irrelevant information, i.e., the background color. This indicates that the proposed method can effectively learn the task-relevant information and ignore the task-irrelevant information, which is consistent with the results in previous subsection.}

\subsection{Detection Performance on Semantic-Shifted Data}
\g{In Table\,\ref{table-detection_only}, we present the detection performance with different $\text{PSNR}_{\text{test}}$ on LSUN and ImageNet dataset. The models are all trained on CIFAR10 dataset~\cite{cifar} with $\text{PSNR}_{\text{train}}=20$ dB. It is shown that the proposed method outperforms the two baselines in both dataseets. Furthermore, the proposed method and DeepJSCC-ODIN schemes are more robust to the channel noise than VIB-ODIN scheme. This is because the VIB-ODIN assumes a fixed latent prior~\cite{vfe} for approximation, resulting in the compromised detection performance in low PSNR regime. However, the proposed method can avoid this issue by assigning label-dependent prior and thus can achieve a better detection performance.}

\subsection{Unified Performance of Generalization and Detection}

\g{In Table\,\ref{tab:joint_ood}, we present the results of joint generalization and detection performance when the testing data contains both domian-shifted and semantic-shifted data. 
The generalization performance is evaluated on Colored-MNIST and Colored-Object dataset, respectively.
The detection performance is reported as the AUROC on LSUN/ImageNet dataset. 
We consider all baselines for comparison, including detection-related methods and generalization-related methods.
For the proposed method, we show the performance of VIFE+VLFE (objective~\eqref{eq:final} for joint detection and generalization) and VIFE (objective~\eqref{eq:vife} for generalization only). 
It can be observed that the proposed method can achieve the best performance on both generalization and detection tasks. Moreover, the generalization performance is slightly compromised when adding the detection objective, which indicates that there is a tradeoff between generalization and detection performance in the proposed method.}

\subsection{Ablation Study}
\g{The hyperparameter selection is a challenging problem  for task-oriented communication without a priori knowledge of the test data.}
In this subsection, we evaluate the robustness of the proposed method with different hyperparameters. We investigate the generalization performance of the proposed method under different $\beta\in\{10^{-5},10^{-4},10^{-3},10^{-2}\}$ and $\lambda\in\{10^3,10^4,10^5,10^6\}$ in dynamic channel conditions for classification tasks over Colored-MNIST and Colored-Object.
As illustrated in Fig.\,\ref{fig:ablation}, the proposed method can achieve a robust performance under a wide spectrum of hyperparameters, i.e., given same $\text{PSNR}_{\text{test}}$, the performance of the proposed method is stable under different hyperparameters. It demonstrates that the proposed method is robust to hyperparameters selection and does not require a fine-tuning process before deployment.

\begin{table}[t]
    \selectfont
    \caption{The detection performance under different testing PSNR on LSUN and ImageNet dataset.}
    \begin{center}
    \scalebox{0.95}{
    \begin{tabular}{@{}c|cccccc@{}}
    \toprule \textbf{LSUN} & 10 dB & 12 dB & 14 dB & 16 dB & 18 dB & 20 dB \\
    \midrule
    DeepJSCC-ODIN & 86.09 & 87.18 & 87.94 & 88.37 & 88.63 & 89.00  \\
    VIB-ODIN & 63.31 & 66.34 & 69.33 & 71.99 & 73.39 & 74.73  \\
    \textbf{Proposed} & \textbf{94.98} & \textbf{95.47} & \textbf{95.93} & \textbf{96.22} & \textbf{96.31} & \textbf{96.42}  \\
    \midrule \textbf{ImageNet} & 10 dB & 12 dB & 14 dB & 16 dB & 18 dB & 20 dB\\
    \midrule
    DeepJSCC-ODIN & 91.90 & 92.14 & 92.74 & 93.06 & 93.37 & 93.57  \\
    VIB-ODIN & 69.03 & 73.50 & 76.90 & 79.78 & 81.66 & 82.39  \\
    \textbf{Proposed} & \textbf{97.98} & \textbf{98.25} & \textbf{98.45} & \textbf{98.48} & \textbf{98.54} & \textbf{98.62} \\
        \bottomrule
    \end{tabular}
    }
    \label{table-detection_only}
    \end{center}
\end{table}
\begin{table}[t]
    \centering
    \caption{Performance of Joint Generalization and Detection}
    \scalebox{0.90}{
    \begin{tabular}{@{}lcc|cc@{}}
    \toprule
    \multirow{2}[2]{*}{\textbf{Method}} & \multicolumn{2}{c}{Colored-MNIST} & \multicolumn{2}{c}{Colored-Object}  \\
    & \textbf{Gen. Acc.}  & $\textbf{AUROC}$  & \textbf{Gen. Acc.}  & $\textbf{AUROC}$ \\
    
    \midrule
    \emph{PSNR= 10dB}\\
    {DeepJSCC-ODIN}     &  N/A    & 76.27/64.07  & N/A  & 71.16/66.41  \\
    {VIB-ODIN}          &  N/A     & 73.21/60.06  & N/A & 72.10/59.56\\
    {DeepJSCC-IRM}      &  62.64  & N/A & 83.19 & N/A\\
    \textbf{VIFE}(ours) &  64.64 & N/A & 85.39 & N/A\\
    \textbf{VIFE+VLFE}(ours)   &  63.9     & 90.72/87.37  & 83.73 & 89.46/85.97 \\
    \midrule
    \emph{PSNR = 20dB}\\
    {DeepJSCC-ODIN}     &  N/A    & 80.12/79.05  & N/A & 75.77/70.34  \\
    {VIB-ODIN}          &  N/A    & 76.40/73.92  & N/A & 74.33/64.32\\
    {DeepJSCC-IRM}      &  65.97      & N/A & 84.06 & N/A\\
    \textbf{VIFE}(ours) &  68.62 & N/A & 87.50 & N/A \\
    \textbf{VIFE+VLFE}(ours)   &  66.07    & 98.92/99.42  & 85.57 & 94.72/91.53 \\   
    \bottomrule
    \end{tabular}%
    }
    \label{tab:joint_ood}%
    \end{table}%

\section{Conclusion}\label{sec-conclusion}
\g{In this work, we investigated the distribution shifts issue in task-oriented communication. Building upon the IB principle and IRM approach, we proposed a novel method to learn the task-relevant information and filter out the task-irrelevant information. Furthermore, a novel method to detect the semantic-shifted data was proposed to enhance the robustness of task-oriented communication. We evaluated the proposed method on Colored-MNIST and Colored-Object classification tasks. The results illustrate that the proposed method can achieve a better generalization and detection performance with distribution shifts. Furthermore, the proposed method provides a new perspective for the design of task-oriented communication and can be extended in several directions such as causal inference for time-varying data and multi-user communication systems.}

\appendices

\section{Derivation of the weight of the two examples}
\label{appendix_example}
For the data generation and transmission model in~\eqref{eq:data_gen}-\eqref{eq:regression_noise} we denote the Gaussian variables in distribution of $U_C$, $U_S$ and label noise as $U_C\sim N(0,\sigma_{d_1}^2),n\sim N(0,\sigma^2)\text{ and }n_S\sim N(0,\sigma^2_{d_2})$. And the channel noise is described by $\eta\sim N(0,\sigma^2_c)$. In regression task, we assume the coefficients are learned by minimizing MSE loss $\mathcal{L}_{MSE}=\mathbb{E}[(Y-\hat{Y})^2]$.

If we only regress on $\hat{U}_C$, we can obtain the optimal $w_1$ by following while $w_2=0$,
\begin{equation}
    \small
    \begin{aligned}
        w_1&=\argmin_{w_1}\mathbb{E}[((U_C+n)-w_1(U_S+\eta))^2]\\
        &=\sigma^2_{d_1}/(\sigma^2_{d_1}+\sigma^2_c)\approx 1(\text{with sufficient small } \sigma^2_c)
    \end{aligned}
\end{equation}

If we regress on $\hat{U}_C$ and $\hat{U}_S$, to get the optimal corresponding coefficients $(w_1,w_2)$, we first derive $\mathcal{L}_{MSE}$:
\begin{equation}
    \small
    \begin{aligned}
        \mathcal{L}_{MSE}=&\mathbb{E}[((U_C+n)-(w_1\hat{U}_C+w_2\hat{U}_S))^2]\\
        =&(1-w_1-w_2)^2\sigma^2_{d_1}+w^2_2\sigma^2_{d_2}\\
        &+(w^2_1+w^2_2)\sigma^2_c+(w_2-1)^2\sigma^2.
    \end{aligned}
\end{equation}
By solving $\frac{\partial \mathcal{L}_{MSE}}{\partial w_1}=0, \frac{\partial \mathcal{L}_{MSE}}{\partial w_2}=0$, we have
\begin{equation}
    \small
    \begin{aligned}
    w_1 &=\frac{2\sigma^2_{d_1}\sigma^2_{d_2}+\sigma^2_c\sigma^2_{d_1}}{\sigma^2_{d_1}(2\sigma^2_{d_2}+\sigma^2)+\sigma^2_c(2\sigma^2_{d_1}+2\sigma^2_{d_2}+\sigma^2_c+\sigma^2)}\\
    &\approx \frac{2\sigma^2_{d_2}}{2\sigma^2_{d_2}+\sigma^2}(\text{with sufficient small } \sigma^2_c)\\
    w_2 &=\frac{\sigma^2_{d_1}\sigma^2+\sigma^2_c(\sigma^2_{d_1}+\sigma^2)}{\sigma^2_{d_1}(2\sigma^2_{d_2}+\sigma^2)+\sigma^2_c(2\sigma^2_{d_1}+2\sigma^2_{d_2}+\sigma^2_c+\sigma^2)}\\
    &\approx \frac{\sigma^2}{2\sigma^2_{d_2}+\sigma^2}(\text{with sufficient small } \sigma^2_c)
    \end{aligned}
\end{equation}

\section{Derivation of the vartional upper bound}
\label{app:vlfe}
According to the non-negativity of Kullback-Leibler divergence, if we use the variational distribution $q(y|\mathbf{\hat{z}})$, $r(\mathbf{\hat{z}}|y)$ to approximate $p(y|\mathbf{\hat{z}})$ and the true conditional latent prior $p(\mathbf{\hat{z}}|y)$, we can derive the variational upper bound of $\mathcal{L}_{VLFE}$ in \eqref{eq:proof}.
\begin{figure*}
\begin{equation}\label{eq:proof}
    \scriptstyle
    \small
        \begin{aligned}
            \mathcal{L}_{LFE}=& \beta\int p(\mathbf{x},y,\mathbf{\hat{z}})\log \frac{p(\mathbf{x},\mathbf{\hat{z}}|y)}{p(\mathbf{x}|y)p(\mathbf{\hat{z}}|y)}d\mathbf{x}dyd\mathbf{\hat{z}}-\int p(\mathbf{x},y,\mathbf{\hat{z}})\log\frac{p(y,\mathbf{\hat{z}})}{p(y)p(\mathbf{\hat{z}})}d\mathbf{x}dyd\mathbf{\hat{z}}\\
            \leq& \beta\int p(\mathbf{x},y)p_{\theta}(\mathbf{\hat{z}}|\mathbf{x})\log\frac{p_{\theta}(\mathbf{\hat{z}}|\mathbf{x})}{r(\mathbf{\hat{z}}|y)}     d\mathbf{x}dyd\mathbf{\hat{z}}-\int   p(\mathbf{x},y)p_{\theta}(\mathbf{\hat{z}}|\mathbf{x})\log q_{\phi}(y|\mathbf{\hat{z}})  d\mathbf{x}dyd\mathbf{\hat{z}}-\underbrace{ H(Y)}_{constant}\\
            =&\beta\int p(\mathbf{x},y)D_{KL}(p_{\theta}(\mathbf{\hat{z}}|\mathbf{x})||r(\mathbf{\hat{z}}|y))d\mathbf{x}dyd\mathbf{\hat{z}}-\int   p(\mathbf{x},y)p_{\theta}(\mathbf{\hat{z}}|\mathbf{x})\log q_{\phi}(y|\mathbf{\hat{z}})  d\mathbf{x}dyd\mathbf{\hat{z}}+\text{constant}\\
            \triangleq& \mathcal{L}_{VLFE}
        \end{aligned}
    \end{equation}
    \hrule
\end{figure*}

\linespread{0.88}{
\bibliographystyle{./bibtex/IEEEtran}
\bibliography{ref}
}
\end{document}